\documentclass[12pt]{article}%
\usepackage{amsmath}
\usepackage{graphicx}
\usepackage{amsfonts}
\usepackage{amssymb}%
\providecommand{\U}[1]{\protect\rule{.1in}{.1in}}
\providecommand{\U}[1]{\protect \rule{.1in}{.1in}}
\providecommand{\U}[1]{\protect \rule{.1in}{.1in}}

\newtheorem{theorem}{Theorem}

\newtheorem{corollary}{Corollary}

\newtheorem{definition}{Definition}

\newtheorem{lemma}{Lemma}

\newtheorem{proposition}{Proposition}

\begin{document}

\title{Conditional divergence risk measures}
\author{Giulio Principi$^{a,}$\thanks{\bigskip\textit{email address: }gp2187@nyu.edu}
\ and Fabio Maccheroni$^{b,}$\thanks{\bigskip\textit{email address:
}fabio.maccheroni@unibocconi.it}\\$^{a}${\small New York University, }$^{b}${\small Universit\`{a} Bocconi and
IGIER}}
\date{November\ 2022}
\maketitle

\begin{abstract}
Our paper contributes to the theory of conditional risk measures and
conditional certainty equivalents. We adopt a random modular approach which
proved to be effective in the study of modular convex analysis and conditional
risk measures. In particular, we study the conditional counterpart of
optimized certainty equivalents. In the process, we provide representation
results for niveloids in the conditional $L^{\infty}$-space. By employing such
representation results we retrieve a conditional version of the variational
formula for optimized certainty equivalents. In conclusion, we apply this
formula to provide a variational representation of the conditional entropic
risk measure.


\end{abstract}

\section{Introduction}

In recent years, growing attention has been devoted to the study of
\textit{conditional} finance. Consider a setting with two time spots $0$ and
$T$. At time $0,$ the risk associated with a certain asset must be assessed,
and at time $T$ the asset pays its final payoff. In the study of unconditional
risk measures, one of the main assumptions is that these measures map random
variables to real numbers. This assumption requires the information available
at time $0$ to be \textit{trivial}. If instead we allow for the possibility of
additional information at time $0$, then a risk measure would assign to each
asset a random variable consistent with the information available at time $0$.
Such random variable represents the risk assessment performed by the
institution consistently with the information available at the time of the
evaluation. There have been several contributions to address the question on
how to generalize dual representations for real valued risk measures to the
conditional setting (see, for instance, Bion-Nadal \cite{BionNad}, Detlefsen
and Scandolo \cite{DefScand}, Filipovi\'{c}, Kupper, and Vogelpoth
\cite{AppCondRisk}, Frittelli and Maggis \cite{FriMaggisCompleteDua},
\cite{FriMaggiModules}, Guo, Zhao, and Zeng \cite{Guo}). As presented by
\cite{AppCondRisk}, two settings have been proposed to study conditional risk
measures. One setting relies upon convex analytic techniques on Lebesgue
spaces, while the other is based on random modular convex analysis (see
\cite{Guo}). In this work, we adopt the latter.

Following the seminal work of Frittelli and Maggis \cite{FriMaggi}, we focus
on optimized certainty equivalents in a conditional framework. In this
setting, we retrieve the optimized certainty equivalent (OCE) representation
for $\phi$-divergence risk measures. The OCE is a decision theoretic criterion
introduced by Ben-Tal and Teboulle \cite{BenTalTeb3}, who, in their subsequent
research, show how this concept includes a wide family of risk measures and
admits a variational representation (see Theorem 4.2 in Ben-Tal and Teboulle
\cite{BenTalTebEq}). This representation highlights how optimized certainty
equivalents emerge from a variational principle based on $\phi$-divergences.
Our main contribution consists in providing a conditional version of this
variational formula by using the modular counterpart of the generalized
Donsker-Varadhan formula (see for instance Dupuis and Ellis \cite{DupEll}) and
representation results for monotone and translation invariant operators
mapping random variables into random variables. In particular, we study
niveloids (Dolecki and Greco \cite{DolGre} and Cerreia-Vioglio et al.
\cite{NiveCerreia}) in the setting of random modules focusing on
representation results for the smallest niveloid dominating an $L^{0}$-concave
operator. The mathematical setting we adopt is close to the one analyzed by
Filipovi\'{c}, Kupper, and Vogelpoth \cite{SepandDual}.

The paper is organized as follows. Section 2 introduces mathematical
preliminaries related to random modules following Cerreia-Vioglio et al.
\cite{SimKupMacVog} and \cite{SepandDual}. Section 3 is the mathematical core
of our work. Here, the representation results for the smallest niveloid are
formally stated and discussed. Section 4 is devoted to applying results from
Section 3 to retrieve the OCE representation. Sections 5, 6, and 7 feature the
concluding section, the acknowledgments, and the Appendix, respectively. In
the Appendix, the reader can find the proofs for the results in the main text
and additional lemmas.

\section{Conditional $L^{p}$-spaces}

Let $(S,\mathcal{F},\mu)$ be a complete probability space and $\mathcal{G}%
\subseteq\mathcal{F}$ be a complete sigma-subalgebra. Here, and hereafter, we
maintain the usual convention to identify measurable functions with their
equivalence classes, ordered with respect to the almost sure pointwise
dominance. We denote by $\bar{L}^{0}(\mathcal{G})$ the set of $\mathcal{G}%
$-measurable extended valued random variables, by $L^{0}(\mathcal{G})$ the set
of $\mathcal{G}$-measurable real valued random variables, by $\bar{L}%
^{0}(\mathcal{G})_{+}$ and $\bar{L}^{0}(\mathcal{G})_{++},$ respectively, the
sets of positive and strictly positive elements of $\bar{L}^{0}(\mathcal{G}).$
The conditional expectation operator $\mathbb{E}_{\mu}[\cdot|\mathcal{G}%
]:L^{1}(\mathcal{F})\rightarrow L^{1}(\mathcal{G})$ is commonly defined as a
function which satisfies the following%
\[
\int_{A}xd\mu=\int_{A}\mathbb{E}_{\mu}[x|\mathcal{G}]d\mu
\]
for all $A\in\mathcal{G}$ and $x\in L^{1}(\mathcal{F}).$ If we restrict to
positive random variables, the conditional expectation operator admits the
following extension%
\begin{align*}
\mathbb{\tilde{E}}_{\mu}[\cdot|\mathcal{G}]  &  :L^{0}(\mathcal{F}%
)_{+}\rightarrow\bar{L}^{0}(\mathcal{G})_{+}\\
x  &  \mapsto\lim_{n\rightarrow\infty}\mathbb{E}_{\mu}[x\wedge n|\mathcal{G}].
\end{align*}
Given this extension, we can define the function $||\cdot||_{p}^{\mathcal{G}%
}:L^{0}(\mathcal{F})\rightarrow\bar{L}^{0}(\mathcal{G})_{+}$ as follows%
\[
||\cdot||_{p}^{\mathcal{G}}:x\mapsto\left\{
\begin{array}
[c]{cc}%
\mathbb{\tilde{E}}_{\mu}[|x|^{p}|\mathcal{G}]^{\frac{1}{p}} & \text{\emph{if
}}p\in\lbrack1,\infty)\\
\inf\{a\in\bar{L}^{0}(\mathcal{G})_{+}:|x|\leq a\} & \text{\emph{if }}p=\infty
\end{array}
\right.
\]
and the conditional counterpart of $L^{p}$-spaces,%
\[
L^{p}(\mathcal{F}|\mathcal{G})=\left\{  x\in L^{0}(\mathcal{F}):||x||_{p}%
^{\mathcal{G}}\in L^{0}(\mathcal{G})\right\}  .
\]
The function $||\cdot||_{p}^{\mathcal{G}}$ satisfies the following properties

\begin{enumerate}
\item $||x||_{p}^{\mathcal{G}}=\mathbf{0}_{S}$ if and only if $x=\mathbf{0}%
_{S},$

\item $||ax||_{p}^{\mathcal{G}}=|a|$ $||x||_{p}^{\mathcal{G}}$ for all $x\in
L^{p}(\mathcal{F}|\mathcal{G})$ and all $a\in L^{0}(\mathcal{G}),$

\item $||x+y||_{p}^{\mathcal{G}}\leq||x||_{p}^{\mathcal{G}}+||y||_{p}%
^{\mathcal{G}}$ for all $x,y\in L^{p}(\mathcal{F}|\mathcal{G}).$
\end{enumerate}

Therefore, $||\cdot||_{p}^{\mathcal{G}}$ is termed $L^{0}(\mathcal{G})$-$p$
norm and it induces a module topology over $L^{p}(\mathcal{F}|\mathcal{G})$.
Moreover, we endow $L^{0}(\mathcal{G})$ with the ring (or strong order)
topology induced by $|\cdot|:L^{0}(\mathcal{G})\rightarrow L^{0}%
(\mathcal{G}),$ which is the one generated by the following base at
$\mathbf{0}_{S},$%
\[
\{B_{\varepsilon}^{\mathcal{G}}:\varepsilon\in L^{0}(\mathcal{G})_{++}\}
\]
where%
\[
B_{\varepsilon}^{\mathcal{G}}=\{x\in L^{0}(\mathcal{G}):|x|\leq\varepsilon\}.
\]
This topology is finer than the topology of convergence in probability (see
\cite{SepandDual} for further details). The general version of the conditional
expectation operator $\mathbb{E}_{\mu}[\cdot|\mathcal{G}]:L^{p}(\mathcal{F}%
|\mathcal{G})\rightarrow L^{0}(\mathcal{G})$ is defined as
\[
\mathbb{E}_{\mu}[x|\mathcal{G}]=\mathbb{\tilde{E}}_{\mu}[x^{+}|\mathcal{G}%
]-\mathbb{\tilde{E}}_{\mu}[x^{-}|\mathcal{G}].
\]
for all $x\in L^{p}(\mathcal{F}|\mathcal{G})$ and all $p\in\lbrack1,\infty]$.
We say that $(L^{p}(\mathcal{F}|\mathcal{G}),||\cdot||_{p}^{\mathcal{G}})$
forms an $L^{0}$-normed module for all $p\in\lbrack1,\infty]$. We adopt the
usual convention $\mathbf{0}_{S}(\pm\infty)=\mathbf{0}_{S}$ and conclude this
introductory section with some definitions.

\begin{definition}
Let $f:L^{p}(\mathcal{F}|\mathcal{G})\rightarrow\bar{L}^{0}(\mathcal{G})$ and
$p\in\lbrack1,\infty]$, we say that $f$ is proper if $f(x)<\infty$ for each
$x\in L^{p}(\mathcal{F}|\mathcal{G})$ and $f(x_{0})>-\infty$ for some
$x_{0}\in L^{p}(\mathcal{F}|\mathcal{G}).$ Suppose $f$ is proper, we say that

\begin{enumerate}
\item $f$ is $L^{0}(\mathcal{G})$-convex if
\[
f(ax+(\mathbf{1}_{S}-a)y)\leq af(x)+(\mathbf{1}_{S}-a)f(y)
\]
for all $a\in L^{0}(\mathcal{G})$ with $\mathbf{0}_{S}\leq a\leq\mathbf{1}%
_{S}$ and all $x,y\in L^{p}(\mathcal{F}|\mathcal{G}).$

\item $f$ is $L^{0}(\mathcal{G})$-concave if
\[
f(ax+(\mathbf{1}_{S}-a)y)\geq af(x)+(\mathbf{1}_{S}-a)f(y)
\]
for all $a\in L^{0}(\mathcal{G})$ with $\mathbf{0}_{S}\leq a\leq\mathbf{1}%
_{S}$ and all $x,y\in L^{p}(\mathcal{F}|\mathcal{G}).$

\item $f$ is $L^{0}(\mathcal{G})$-translation invariant if
\[
f(x+a)=f(x)+a
\]
for all $a\in L^{0}(\mathcal{G})$ and all $x\in L^{p}(\mathcal{F}%
|\mathcal{G}).$

\item $f$ is $L^{0}(\mathcal{G})$-linear if
\[
f(ax+by)=af(x)+bf(y)
\]
for all $a,b\in L^{0}(\mathcal{G})$ and all $x,y\in L^{p}(\mathcal{F}%
|\mathcal{G}).$

\item $f$ is monotone if $f(x)\geq f(y)$ whenever $x\geq y$ for all $x,y\in
L^{p}(\mathcal{F}|\mathcal{G}).$

\item $f$ is a $L^{0}(\mathcal{G})$-niveloid if it is monotone and
$L^{0}(\mathcal{G})$-translation invariant.

\item For each topology $\tau$ on $L^{p}(\mathcal{F}|\mathcal{G})$, $f$ is
$(\tau,L^{0}(\mathcal{G}))$-upper semicontinuous if
\[
\left\{  x\in L^{p}(\mathcal{F}|\mathcal{G}):f(x)\geq a\right\}
\]
is $\tau$-closed for all $a\in L^{0}(\mathcal{G}).$
\end{enumerate}
\end{definition}

In what follows we will refer to $(\tau,L^{0}(\mathcal{G}))$-upper
semicontinuity as $\tau$-upper semicontinuity unless differently specified.
Notice that the assumption that the function $f$ is proper, in the definition
above, helps avoiding problems related to extended arithmetic.

\section{Niveloidification in the conditional $L^{\infty}$-space}

Following \cite{SimKupMacVog}, we see that $(L^{\infty}(\mathcal{F}%
|\mathcal{G}),L^{1}(\mathcal{F}|\mathcal{G}))$ is a conditional dual pair.
Indeed, $\left\langle \cdot,\cdot\right\rangle ^{\mathcal{G}}:(x,y)\mapsto
\mathbb{E}_{\mu}[xy|\mathcal{G}]$ is such that%
\[
\mathbb{E}_{\mu}[|xy||\mathcal{G}]\leq||x||_{\infty}^{\mathcal{G}}%
||x||_{1}^{\mathcal{G}}\in L^{0}(\mathcal{G})
\]
for all $(x,y)\in L^{\infty}(\mathcal{F}|\mathcal{G})\times L^{1}%
(\mathcal{F}|\mathcal{G}),$ and both $L^{\infty}(\mathcal{F}|\mathcal{G})$ and
$L^{1}(\mathcal{F}|\mathcal{G})$ contain all the indicator functions
measurable with respect to $\mathcal{F}$. Now, identifying $L^{1}%
(\mathcal{F}|\mathcal{G})$ as a nonempty subset of $L^{0}(\mathcal{G})$-linear
maps over $L^{\infty}(\mathcal{F}|\mathcal{G}),$ we denote by $\sigma
=\sigma(L^{\infty}(\mathcal{F}|\mathcal{G}),L^{1}(\mathcal{F}|\mathcal{G}))$
the weakest topology that makes $L^{0}(\mathcal{G})$-linear maps in
$L^{1}(\mathcal{F}|\mathcal{G})$ continuous. Later on, we will use the
following notation, given a topological space $\left(  L^{\infty}%
(\mathcal{F}|\mathcal{G}),\tau\right)  ,$%
\[
\mathtt{Hom}_{L^{0}(\mathcal{G})}^{\tau}(L^{\infty}(\mathcal{F}|\mathcal{G}%
),L^{0}(\mathcal{G}))=\left\{  f:L^{\infty}(\mathcal{F}|\mathcal{G}%
)\rightarrow L^{0}(\mathcal{G}):f\text{ is }L^{0}(\mathcal{G})\text{-linear
and }\tau\text{-continuous}\right\}
\]
denote the module of $\tau$-continuous module homomorphisms on $L^{\infty
}(\mathcal{F}|\mathcal{G})$.\footnote{To ease the notation we will make use of
the following (small) abuse%
\[
\mathtt{Hom}_{L^{0}(\mathcal{G})}^{\left\Vert \cdot\right\Vert _{\infty
}^{\mathcal{G}}}(L^{\infty}(\mathcal{F}|\mathcal{G}),L^{0}(\mathcal{G}))
\]
\par
to denote the module of $\left\Vert \cdot\right\Vert _{\infty}^{\mathcal{G}}%
$-continuous module homomorphisms on $L^{\infty}(\mathcal{F}|\mathcal{G}).$}
Then, a direct application of Corollary 1 in\ \cite{SimKupMacVog} yields the following.

\begin{lemma}
\label{homosex}If $L^{\infty}(\mathcal{F}|\mathcal{G})$ is endowed with
$\sigma,$ then $L^{1}(\mathcal{F}|\mathcal{G})=\mathtt{Hom}_{L^{0}%
(\mathcal{G})}^{\sigma}(L^{\infty}(\mathcal{F}|\mathcal{G}),L^{0}%
(\mathcal{G})).$ That is, for all $\sigma$-continuous $L^{0}(\mathcal{G}%
)$-linear maps $f:L^{\infty}(\mathcal{F}|\mathcal{G})\rightarrow
L^{0}(\mathcal{G}),$ there exists a unique $y\in L^{1}(\mathcal{F}%
|\mathcal{G})$ such that%
\[
f(x)=\mathbb{E}_{\mu}[xy|\mathcal{G}]
\]
for all $x\in L^{\infty}(\mathcal{F}|\mathcal{G})$. Conversely, $x\mapsto
\mathbb{E}_{\mu}[xy|\mathcal{G}]$ defines a $\sigma$-continuous $L^{0}%
(\mathcal{G})$-linear map on $L^{\infty}(\mathcal{F}|\mathcal{G})$ for all
$y\in L^{1}(\mathcal{F}|\mathcal{G})$.
\end{lemma}

Passing to concave duality, a straightforward application of Lemma
\ref{homosex} above and Theorem 3.13 in Guo, Zhao, and Zeng \cite{Guo} yields
the following representation.

\begin{proposition}
\label{modu-dua-rep}If $I:L^{\infty}(\mathcal{F}|\mathcal{G})\rightarrow
\bar{L}^{0}(\mathcal{G})$\ is $\sigma$-upper semicontinuous, proper, and
$L^{0}(\mathcal{G})$-concave, then%
\[
I(x)=\inf_{y\in L^{1}(\mathcal{F}|\mathcal{G})}\left\{  \mathbb{E}_{\mu
}[xy|\mathcal{G}]+c\left(  y\right)  \right\}
\]
for all $x\in L^{\infty}(\mathcal{F}|\mathcal{G}),$ where $c:L^{1}%
(\mathcal{F}|\mathcal{G})\rightarrow\bar{L}^{0}(\mathcal{G})$ is defined as%
\[
c(y)=\sup_{z\in L^{\infty}(\mathcal{F}|\mathcal{G})}\left\{  I\left(
z\right)  -\mathbb{E}_{\mu}[zy|\mathcal{G}]\right\}
\]
for all $y\in L^{1}(\mathcal{F}|\mathcal{G})$.
\end{proposition}

Hereafter, we will always refer to $c$ as the function defined in Proposition
\ref{modu-dua-rep}. Now suppose that instead of $L^{0}(\mathcal{G})$-concavity
we ask $I$ to be monotone and $L^{0}(\mathcal{G})$-translation invariant. We
have that for all $x,y\in L^{\infty}(\mathcal{F}|\mathcal{G})$ and all $a\in
L^{0}(\mathcal{G}),$ if $a+y\leq x,$ then $a+I(y)\leq I(x).$ Thus,
$\sup_{a+y\leq x}\left\{  a+I\left(  y\right)  \right\}  \leq I(x).$ In
addition, since $a+(x-a)\leq x,$%

\[
I(x)=a+I(x-a)\leq\sup_{a+y\leq x}\left\{  a+I\left(  y\right)  \right\}  .
\]
Therefore, whenever $I$ is monotone and $L^{0}(\mathcal{G})$-translation
invariant, we have $I(x)=\sup_{a+y\leq x}\left\{  a+I\left(  y\right)
\right\}  $ for all $x\in L^{\infty}(\mathcal{F}|\mathcal{G}).$ This simple
remark raises an interesting question: which is the smallest monotone and
$L^{0}(\mathcal{G})$-translation invariant operator dominating some map
$T:L^{\infty}(\mathcal{F}|\mathcal{G})\rightarrow L^{0}(\mathcal{G}%
).$\footnote{This problem was also studied in \cite{AppCondRisk}, our analysis
differs in the setting we adopt and some of the proofs we propose.} The
discussion right above suggests the following candidate
$I_{\operatorname*{niv}}:L^{\infty}(\mathcal{F}|\mathcal{G})\rightarrow\bar
{L}^{0}(\mathcal{G})$ defined as%
\begin{equation}
I_{\operatorname*{niv}}\left(  x\right)  =\sup_{(a,y)\in C(x)}\left\{
a+I\left(  y\right)  \right\}  \label{eq:def-I-niv}%
\end{equation}
where%
\[
C(x)=\left\{  \left(  a,y\right)  \in L^{0}(\mathcal{G})\mathbb{\times
}L^{\infty}(\mathcal{F}|\mathcal{G}):a+y\leq x\right\}
\]
for all $x\in L^{\infty}(\mathcal{F}|\mathcal{G}).$\footnote{Notice that the
definition is well posed since: for all $x\in L^{\infty}(\mathcal{F}%
|\mathcal{G})$ and all $a\in L^{0}(\mathcal{G}),$ $(a,x-a)\in C(x);$ $\bar
{L}^{0}(\mathcal{G})$ satisfies the countable sup property; $\bar{L}%
^{0}(\mathcal{G})$ is a complete lattice with respect to the almost sure
dominance.} \ In what follows we study the properties of this operator. To do
so, we define the following sets,%
\[
\Delta(\mathcal{F}|\mathcal{G})\mathbf{=}\left\{  y\in L^{1}(\mathcal{F}%
|\mathcal{G})_{+}:\mathbb{E}_{\mu}[y|\mathcal{G}]=\mathbf{1}_{S}\right\}
\]
and, for all functions $f:L^{\infty}(\mathcal{F}|\mathcal{G})\rightarrow
\bar{L}^{0}(\mathcal{G}),$%
\[
\operatorname*{dom}f=\left\{  x\in L^{\infty}(\mathcal{F}|\mathcal{G}):f(x)\in
L^{0}(\mathcal{G})\right\}  .
\]

The set $\Delta(\mathcal{F}|\mathcal{G})$ can be seen as the set of
\textit{conditional densities}.\footnote{This set is subtly different from the
following%
\[
Y=\left\{  y\in L^{1}(\mathcal{F}|\mathcal{G})_{+}\text{ }:\int_{S}%
yd\mu=1\right\}  .
\]
\par
Indeed, while we have $\Delta(\mathcal{F}|\mathcal{G})\subseteq Y,$ the
converse inclusion does not hold in general.} In order to provide a
representation for $I_{\operatorname*{niv}}$, we start asking under which
conditions $I_{\operatorname*{niv}}$ is $L^{0}(\mathcal{G})$-valued and which
properties it inherits from $I.$ The next two results answer to these
questions, and provide some conditions that will be relevant to characterize
$I_{\operatorname*{niv}}.$

\begin{proposition}
\label{pro:dua-niv}If $I:L^{\infty}(\mathcal{F}|\mathcal{G})\rightarrow
L^{0}(\mathcal{G})$\ is $\sigma$-upper semicontinuous and $L^{0}(\mathcal{G}%
)$-concave, then%
\begin{equation}
I_{\operatorname*{niv}}\left(  x\right)  \leq\inf_{y\in\Delta(\mathcal{F}%
|\mathcal{G})}\left\{  \mathbb{E}_{\mu}[xy|\mathcal{G}]+c\left(  y\right)
\right\}  \label{eq:dua-niv-rep}%
\end{equation}
for all $x\in L^{\infty}(\mathcal{F}|\mathcal{G}).$ Moreover, if
$\operatorname*{dom}c\cap\Delta(\mathcal{F}|\mathcal{G})\not =\emptyset$, then
$I_{\operatorname*{niv}}$ is $L^{0}(\mathcal{G})$-valued.
\end{proposition}

\begin{lemma}
\label{lem:con-niv}If $I:L^{\infty}(\mathcal{F}|\mathcal{G})\rightarrow
L^{0}(\mathcal{G})$\ is $\sigma$-upper semicontinuous, $L^{0}(\mathcal{G}%
)$-concave, and $\operatorname*{dom}c\cap\Delta(\mathcal{F}|\mathcal{G}%
)\not =\emptyset$,\ then $I_{\operatorname*{niv}}$ is a $L^{0}(\mathcal{G}%
)$-valued, $L^{0}(\mathcal{G})$-concave, $L^{0}(\mathcal{G})$-niveloid such
that $I_{\operatorname*{niv}}\geq I$.
\end{lemma}

We are now ready to provide the following representation result for the
smallest $L^{0}(\mathcal{G})$-niveloid dominating an $L^{0}(\mathcal{G}%
)$-concave operator. An analogous result was proved in a different setting in
\cite{AppCondRisk} (see Proposition 4.2).

\begin{lemma}
\label{lem:niv}If $I:L^{\infty}(\mathcal{F}|\mathcal{G})\rightarrow
L^{0}(\mathcal{G})$\ is $\sigma$-upper semicontinuous, $L^{0}(\mathcal{G}%
)$-concave, and $\operatorname*{dom}c\cap\Delta(\mathcal{F}|\mathcal{G}%
)\not =\emptyset$, then $I_{\operatorname*{niv}}$ is the smallest
$L^{0}(\mathcal{G})$-niveloid dominating $I$\ and%
\begin{equation}
I_{\operatorname*{niv}}\left(  x\right)  =\inf_{y\in\Delta(\mathcal{F}%
|\mathcal{G})}\left\{  \mathbb{E}_{\mu}[xy|\mathcal{G}]+c\left(  y\right)
\right\}  \label{eq:dua-sma-niv}%
\end{equation}
for all $x\in L^{\infty}(\mathcal{F}|\mathcal{G}).$
\end{lemma}

This result highlights that also in the conditional setting monotonicity and
$L^{0}(\mathcal{G})$-translation invariance allow to refine the Fenchel-Moreau
representation. In particular, the minimization can be restricted to the set
of conditional densities in place of the entire dual module, this was firstly
observed in \cite{AppCondRisk} (see Corollary 3.14 and Proposition 4.2) and
\cite{Guo} (see Theorem 4.17). In what follows, we show that the set of
conditional densities can be substituted by a set of probability measures.
This leads to a further representation for $I_{\operatorname*{niv}},$ to which
we dedicate the rest of the section.

Let $\Delta^{\sigma}(\mathcal{F})$ be the set of countably additive
probability measures over $\mathcal{F}$ and define,
\[
\mathcal{M(G)=}\left\{  \rho\in\Delta^{\sigma}(\mathcal{F}){\LARGE :}\exists
z_{\rho}\in\Delta(\mathcal{F}|\mathcal{G})\text{ \textit{s.t}. }\forall
A\in\mathcal{F},\text{ }\rho(A)=\int_{S}\mathbb{E}_{\mu}[\mathbf{1}_{A}%
z_{\rho}|\mathcal{G}]d\mu\right\}  .
\]
The following lemma holds.

\begin{lemma}
\label{measrep}Let $y\in\Delta(\mathcal{F}|\mathcal{G}),$ then there exists
$\nu\in\mathcal{M(G)}$ such that%
\[
\mathbb{E}_{\mu}[xy|\mathcal{G}]=\mathbb{E}_{\nu}[x|\mathcal{G}]
\]
for all $x\in L^{\infty}(\mathcal{F}|\mathcal{G}).$ Moreover, the mapping
$T:\mathcal{M(G)}\rightarrow\Delta(\mathcal{F}|\mathcal{G}),$ defined as%
\[
T(\nu)=y_{\nu}%
\]
where $y_{\nu}$ is the representative element of $\nu$ for all $\nu
\in\mathcal{M(G)}$, is a bijection.
\end{lemma}

This lemma guarantees that for each $y\in\Delta(\mathcal{F}|\mathcal{G}),$
there exists a unique $\nu\in\mathcal{M(G)}$ such that%
\[
\mathbb{E}_{\mu}[xy|\mathcal{G}]=\mathbb{E}_{\nu}[x|\mathcal{G}]
\]
for all $x\in L^{\infty}(\mathcal{F}|\mathcal{G}).$ A simple characterization
of $\mathcal{M(G)}$ is the following.

\begin{lemma}
\label{charactM}Let%
\[
\widehat{\mathcal{M}}\mathcal{(G)}=\left\{  \rho\in\Delta^{\sigma}%
(\mathcal{F}):\rho|_{\mathcal{G}}=\mu|_{\mathcal{G}},\text{ }\exists z_{\rho
}\in Y\text{ s.t. }\forall A\in\mathcal{F},\text{ }\rho(A)=\int_{A}z_{\rho
}d\mu\right\}
\]
with $Y=\left\{  y\in L^{1}(\mathcal{F}|\mathcal{G})_{+}:\int_{S}%
yd\mu=1\right\}  .$ Then, $\mathcal{M(G)=}\widehat{\mathcal{M}}\mathcal{(G)}.$
\end{lemma}

This lemma highlights that $\mathcal{M(G)}$ is exactly the set of countably
additive probability measures over $\mathcal{F}$ that are absolutely
continuous with respect to $\mu$ and coincide with $\mu$ over $\mathcal{G}$.
Now we can retrieve the following representation.

\begin{proposition}
\label{measrepniv}If $I:L^{\infty}(\mathcal{F}|\mathcal{G})\rightarrow
L^{0}(\mathcal{G})$\ is $\sigma$-upper semicontinuous, $L^{0}(\mathcal{G}%
)$-concave, and $\operatorname*{dom}c\cap\Delta(\mathcal{F}|\mathcal{G}%
)\not =\emptyset$,\ then%
\begin{equation}
I_{\operatorname*{niv}}\left(  x\right)  =\inf_{\nu\in\mathcal{M(G)}}\left\{
\mathbb{E}_{\nu}[x|\mathcal{G}]+\hat{c}\left(  \nu\right)  \right\}
\end{equation}
for all $x\in L^{\infty}(\mathcal{F}|\mathcal{G}),$ where $\hat{c}%
:\mathcal{M(G)\rightarrow}\bar{L}^{0}(\mathcal{G})$ is defined as%
\[
\hat{c}(\nu)=\sup_{z\in L^{\infty}(\mathcal{F}|\mathcal{G})}\left\{
I(z)-\mathbb{E}_{\nu}[z|\mathcal{G}]\right\}
\]
for all $\nu\in\mathcal{M(G)}$.
\end{proposition}

Therefore, so far we proved the following equality%
\begin{equation}
\inf_{\nu\in\mathcal{M(G)}}\left\{  \mathbb{E}_{\nu}[x|\mathcal{G}]+\hat
{c}\left(  \nu\right)  \right\}  =\sup_{(a,y)\in C(x)}\left\{  a+I\left(
y\right)  \right\}  \label{generalBTT}%
\end{equation}

for all $\sigma$-upper semicontinuous $L^{0}(\mathcal{G})$-concave functions
$I:L^{\infty}(\mathcal{F}|\mathcal{G})\rightarrow L^{0}(\mathcal{G})$ with
$\operatorname*{dom}c\cap\Delta(\mathcal{F}|\mathcal{G})\not =\emptyset$ and
all $x\in L^{\infty}(\mathcal{F}|\mathcal{G}).$ This equality can be seen as a
generalization of the variational representation for optimized certainty
equivalents proved in \cite{BenTalTebEq} (see Theorem 4.2). In the next
section, we provide the actual extension to the conditional $L^{\infty}$-space
of the OCE representation using $\phi$-divergences. To conclude this section,
we state a result which highlights how the order of \textit{niveloidification
}is irrelevant, that is, it does not matter whether we make $I$ monotone
before of making it $L^{0}(\mathcal{G})$-translation invariant or the other
way around.

\begin{proposition}
\label{orderofniv}If $I:L^{\infty}(\mathcal{F}|\mathcal{G})\rightarrow
L^{0}(\mathcal{G})$, then $I_{\operatorname*{niv}}$ is such that%
\[
I_{\operatorname*{niv}}\left(  x\right)  =\sup_{\left\{  y\in L^{\infty
}(\mathcal{F}|\mathcal{G}):y\leq x\right\}  }\sup_{a\in L^{0}(\mathcal{G}%
)}\left\{  a+I\left(  y-a\right)  \right\}
\]
for all $x\in L^{\infty}(\mathcal{F}|\mathcal{G}).$
\end{proposition}

\section{A variational formula for conditional OCEs}

We denote the class of continuous and strictly convex functions $\phi:\left[
0,\infty\right)  \rightarrow\left[  0,\infty\right)  $ with the properties
$\phi\left(  1\right)  =0$ and $\lim_{t\rightarrow\infty}\phi\left(  t\right)
/t=\infty$ by $\Phi$. If $\phi\in\Phi$, then we define by $C\left(
\phi\right)  $ the collection of all lower semicontinuous and
convex\ functions $\hat{\phi}:\mathbb{R}\rightarrow\left(  -\infty
,\infty\right]  $ such that, for all $t\geq0,$%
\[
\phi\left(  t\right)  =\sup_{m\in\mathbb{R}}\left\{  mt-\hat{\phi}\left(
m\right)  \right\}  .
\]
Note that $C\left(  \phi\right)  $ is never empty. In fact, it contains the
function $\phi^{\ast}:\mathbb{R}\rightarrow\left(  -\infty,\infty\right]  $
defined by%
\begin{equation}
\phi^{\ast}:m\mapsto\sup_{t\in\left[  0,\infty\right)  }\left\{
mt-\phi\left(  t\right)  \right\}  .
\end{equation}
In this case, given the assumptions on $\phi$, $\phi^{\ast}$ is real valued,
increasing and the convex conjugate of the lower semicontinuous and convex
extension $\tilde{\phi}:\mathbb{R}\rightarrow\left(  -\infty,\infty\right]  $
of $\phi$ defined by%
\[
\tilde{\phi}:t\mapsto\left\{
\begin{array}
[c]{cc}%
\phi\left(  t\right)  & t\geq0\\
+\infty & t<0
\end{array}
\right.
\]
Given $\hat{\phi}\in C\left(  \phi\right)  $, note that $\hat{\phi}^{\ast}$,
defined by $\hat{\phi}^{\ast}\left(  t\right)  =\sup_{m\in\mathbb{R}}\left\{
mt-\hat{\phi}\left(  m\right)  \right\}  $ for all $t\in\mathbb{R}$, is a
lower semicontinuous,\ and convex extension of $\phi$. Now fix some $\phi
\in\Phi$ and define the operator $I_{\phi}:L^{\infty}(\mathcal{F}%
|\mathcal{G})\rightarrow\bar{L}^{0}(\mathcal{G})$ as%
\[
I_{\phi}(x)=-\mathbb{E}_{\mu}[\phi^{\ast}\left(  -x\right)  |\mathcal{G}]
\]
for all $x\in L^{\infty}(\mathcal{F}|\mathcal{G}).$ We study the properties of
the smallest $L^{0}(\mathcal{G})$-niveloid dominating $I_{\phi}$. In
particular, we apply the results presented in the previous section focusing on
the operator $I_{\phi}$. Interestingly, we prove that our equation
(\ref{generalBTT}) when applied to $I_{\phi}$ yields that variational formula
for the OCE in the conditional setting (see Theorem \ref{Rep_COCEs} below). To
this end we start presenting the properties of $I_{\phi}$. More specifically,
notice that $I_{\phi}$ is $L^{0}(\mathcal{G})$-concave, monotone,
$L^{0}(\mathcal{G})$-valued, and $\sigma$-upper semicontinuous (see Lemma
\ref{PropertiesIphi} in the Appendix). Moreover, we define the following
operator $D_{\phi,\mathcal{G}}\left(  \cdot||\mu\right)
:\mathcal{M(G)\rightarrow}\bar{L}^{0}(\mathcal{G})$ as%
\[
D_{\phi,\mathcal{G}}\left(  \nu||\mu\right)  =\mathbb{E}_{\mu}\left[
\phi\left(  \frac{d\nu}{d\mu}\right)  {\LARGE |}\mathcal{G}\right]
\]
for all $\nu\in\mathcal{M(G)}.$ It is immediate to see that $D_{\phi
,\mathcal{G}}\left(  \cdot||\mu\right)  $ is the conditional version of the
classical $\phi$-divergence. Let $\Delta^{\sigma}\left(  \mu\right)  $ denote
the set of probability measures on $\mathcal{F}$ which are absolutely
continuous with respect to $\mu.$ Notice that $\mathcal{M(G)}\subseteq
\Delta^{\sigma}\left(  \mu\right)  ,$ indeed, if $A\in\mathcal{F}$ is an $\mu
$-null set and $\nu\in\mathcal{M(G)}$, then, for some $y\in\Delta
(\mathcal{F}|\mathcal{G})$%
\[
\nu(A)=\int_{S}\mathbb{E}_{\mu}[\mathbf{1}_{A}y|\mathcal{G}]d\mu=\int
_{S}\mathbf{1}_{A}yd\mu=0
\]
thus $\nu\ll\mu.$ This guarantees that $D_{\phi,\mathcal{G}}\left(  \cdot
||\mu\right)  $ is a well defined function. It is interesting to notice that
also in this conditional setting $D_{\phi,\mathcal{G}}\left(  \cdot
||\mu\right)  $ can be written as the \textit{penalty} \textit{function} of
$I_{\phi}$. In particular, by Theorem \ref{thm:CondiRocky} in the Appendix, we
have that for all $\nu\in\mathcal{M(G)},$%
\begin{align*}
D_{\phi,\mathcal{G}}\left(  \nu||\mu\right)   &  =\sup_{z\in L^{\infty
}(\mathcal{F}|\mathcal{G})}\left\{  \mathbb{E}_{\nu}[z|\mathcal{G}%
]-\mathbb{E}_{\mu}[\phi^{\ast}\left(  z\right)  |\mathcal{G}]\right\} \\
&  =\sup_{z\in L^{\infty}(\mathcal{F}|\mathcal{G})}\left\{  \mathbb{E}_{\nu
}[-z|\mathcal{G}]-\mathbb{E}_{\mu}[\phi^{\ast}\left(  -z\right)
|\mathcal{G}]\right\} \\
&  =\sup_{z\in L^{\infty}(\mathcal{F}|\mathcal{G})}\left\{  I_{\phi
}(z)-\mathbb{E}_{\nu}[z|\mathcal{G}]\right\} \\
&  =:\hat{c}_{\phi}\left(  \nu\right)  .
\end{align*}
This can be seen as a conditional version of the generalized Donsker-Varadhan
formula. Thanks to equation (\ref{generalBTT}) and the equality we just
proved, we are now ready to prove our main result. Since $I_{\phi}$ is
$L^{0}(\mathcal{G})$-valued, $L^{0}(\mathcal{G})$-concave, $\sigma$-upper
semicontinuous, and $\mu\in\operatorname*{dom}D_{\phi,\mathcal{G}}\left(
\cdot||\mu\right)  \cap\mathcal{M(G)}\not =\emptyset$, by Lemma \ref{lem:niv}
and Lemma \ref{measrep} we have that%
\begin{align*}
\inf_{\nu\in\mathcal{M(G)}}\left\{  \mathbb{E}_{\nu}[x|\mathcal{G}%
]+D_{\phi,\mathcal{G}}\left(  \nu||\mu\right)  \right\}   &  =\inf_{\nu
\in\mathcal{M(G)}}\left\{  \mathbb{E}_{\nu}[x|\mathcal{G}]+\hat{c}_{\phi
}\left(  \nu\right)  \right\} \\
&  =I_{\phi,\text{niv}}(x)\\
&  =\sup_{\left(  a,y\right)  \in C\left(  x\right)  }\left\{  a+I_{\phi
}\left(  y\right)  \right\} \\
&  =\sup_{a\in L^{0}(\mathcal{G})}\left\{  a+I_{\phi}\left(  x-a\right)
\right\} \\
&  =\sup_{a\in L^{0}(\mathcal{G})}\left\{  a-\mathbb{E}_{\mu}[\phi^{\ast
}\left(  a-x\right)  |\mathcal{G}]\right\}
\end{align*}
for all $x\in L^{\infty}(\mathcal{F}|\mathcal{G})$. This proves our main
theorem, stated below formally.

\begin{theorem}
\label{Rep_COCEs}For all $x\in L^{\infty}(\mathcal{F}|\mathcal{G})$ and all
$\phi\in\Phi,$%
\[
\inf_{\nu\in\mathcal{M(G)}}\left\{  \mathbb{E}_{\nu}[x|\mathcal{G}%
]+D_{\phi,\mathcal{G}}\left(  \nu||\mu\right)  \right\}  =\sup_{a\in
L^{0}(\mathcal{G})}\left\{  a-\mathbb{E}_{\mu}[\phi^{\ast}\left(  a-x\right)
|\mathcal{G}]\right\}  .
\]

\end{theorem}

Theorem \ref{Rep_COCEs} provides the conditional counterpart to the
variational representation of conditional optimized certainty equivalents.
Some interesting examples can be retrieved from this representation. In
particular, the following corollary is an immediate consequence of Theorem
\ref{Rep_COCEs}. Denoting by $D_{\text{\textit{KL}},\mathcal{G}}\left(
\nu||\mu\right)  $ the conditional Kullback-Leibler divergence between $\nu$
and $\mu$, we have the following.

\begin{corollary}
\label{Rep_entropic_coce} Let $\phi\in\Phi$ be $\phi:x\mapsto x\log(x)-x+1,$
then
\[
\inf_{\nu\in\mathcal{M(G)}}\left\{  \mathbb{E}_{\nu}[x|\mathcal{G}%
]+D_{\text{KL},\mathcal{G}}\left(  \nu||\mu\right)  \right\}  =-\log
\mathbb{E}_{\mu}[e^{-x}|\mathcal{G}]
\]
for all $x\in L^{\infty}(\mathcal{F}|\mathcal{G})$.
\end{corollary}

\section{Conclusion}

In conclusion, this paper provides a first step towards the study of
conditional divergence risk measures. In particular, a conditional version for
the variational representation of optimized certainty equivalents is provided.
Following this line of research, our methodology leads to a deeper
understanding of conditional divergence based risk measures (e.g., entropic
risk measures) and their relation with OCE. The relation between the measure
of information and this decision theoretic concept is particularly
interesting, and to us it appears even more significant in the conditional
setting where the role of additional information is crucial.

\section{Acknowledgments}

We thank Andrea Aveni, Yuval Peres, and the participants of the Weekly
Seminars on Risk Management and Actuarial Science for useful comments. We are
particularly indebted to Simone Cerreia-Vioglio and Ruodu Wang for many
helpful discussions. We thank for the financial support the Ministero
dell'universit\`{a} e della ricerca (grant 2017CY2NCA), and the MacCracken Fellowship.

\section{Appendix}

\subsection{Proofs of the results in the main text}

\noindent\textbf{Proof of Proposition \ref{modu-dua-rep} }First of all, since
$I$ is assumed to be $\sigma$-upper semicontinuous and $L^{0}(\mathcal{G}%
)$-concave, we have that $-I$ is $\sigma$-lower semicontinuous and
$L^{0}(\mathcal{G})$-convex. Moreover, since $I$ is proper, $-I(x)>-\infty$
for all $x\in L^{\infty}(\mathcal{F}|\mathcal{G}),$ and there exists $x_{0}\in
L^{\infty}(\mathcal{F}|\mathcal{G})$ such that $-I(x_{0})<\infty.$ Then, by
Lemma \ref{homosex} and since $\left(  L^{\infty}(\mathcal{F}|\mathcal{G}%
),\sigma\right)  $ is a random locally convex $L^{0}(\mathcal{G})$-module with
$L^{\infty}(\mathcal{F}|\mathcal{G})$ satisfying the countable concatenation
property,\footnote{See [17] and section 7.3 below for further details.}
Theorem 3.13 in \cite{Guo} implies that%
\begin{align*}
-I(x)  &  =(-I)^{\ast\ast}(x)=\sup_{w\in L^{1}(\mathcal{F}|\mathcal{G}%
)}\left\{  \mathbb{E}_{\mu}[xw|\mathcal{G}]-(-I)^{\ast}\left(  w\right)
\right\} \\
&  =\sup_{w\in L^{1}(\mathcal{F}|\mathcal{G})}\left\{  \mathbb{E}_{\mu
}[xw|\mathcal{G}]-\sup_{z\in L^{\infty}(\mathcal{F}|\mathcal{G})}\left\{
\mathbb{E}_{\mu}[zw|\mathcal{G}]+I(z)\right\}  \right\} \\
&  =-\inf_{w\in L^{1}(\mathcal{F}|\mathcal{G})}\left\{  \sup_{z\in L^{\infty
}(\mathcal{F}|\mathcal{G})}\left\{  \mathbb{E}_{\mu}[zw|\mathcal{G}%
]+I(z)\right\}  -\mathbb{E}_{\mu}[xw|\mathcal{G}]\right\} \\
&  =-\inf_{-w\in L^{1}(\mathcal{F}|\mathcal{G})}\left\{  \sup_{z\in L^{\infty
}(\mathcal{F}|\mathcal{G})}\left\{  I(z)-\mathbb{E}_{\mu}[zw|\mathcal{G}%
]\right\}  +\mathbb{E}_{\mu}[xw|\mathcal{G}]\right\} \\
&  =-\inf_{y\in L^{1}(\mathcal{F}|\mathcal{G})}\left\{  \sup_{z\in L^{\infty
}(\mathcal{F}|\mathcal{G})}\left\{  I(z)-\mathbb{E}_{\mu}[zy|\mathcal{G}%
]\right\}  +\mathbb{E}_{\mu}[xy|\mathcal{G}]\right\} \\
&  =-\inf_{y\in L^{1}(\mathcal{F}|\mathcal{G})}\left\{  \mathbb{E}_{\mu
}[xy|\mathcal{G}]+\sup_{z\in L^{\infty}(\mathcal{F}|\mathcal{G})}\left\{
I(z)-\mathbb{E}_{\mu}[zy|\mathcal{G}]\right\}  \right\}
\end{align*}
for all $x\in L^{\infty}(\mathcal{F}|\mathcal{G}).$ Then, the claim follows%
\[
I(x)=\inf_{y\in L^{1}(\mathcal{F}|\mathcal{G})}\left\{  \mathbb{E}_{\mu
}[xy|\mathcal{G}]+c\left(  y\right)  \right\}
\]
for all $x\in L^{\infty}(\mathcal{F}|\mathcal{G}).$\hfill$\blacksquare$

\noindent\textbf{Proof of Proposition \ref{pro:dua-niv} }Fix $x\in L^{\infty
}(\mathcal{F}|\mathcal{G})$. By Proposition \ref{modu-dua-rep} we have that%
\begin{align*}
I_{\operatorname*{niv}}\left(  x\right)   &  =\sup_{\left(  a,y\right)  \in
C\left(  x\right)  }\left\{  a+I\left(  y\right)  \right\}  =\sup_{\left(
a,y\right)  \in C\left(  x\right)  }\left\{  a+\inf_{z\in L^{\infty
}(\mathcal{F}|\mathcal{G})}\left\{  \mathbb{E}_{\mu}[yz|\mathcal{G}]+c\left(
z\right)  \right\}  \right\} \\
&  =\sup_{\left(  a,y\right)  \in C\left(  x\right)  }\inf_{z\in L^{\infty
}(\mathcal{F}|\mathcal{G})}\left\{  a+\mathbb{E}_{\mu}[yz|\mathcal{G}%
]+c\left(  z\right)  \right\} \\
&  \leq\inf_{z\in L^{\infty}(\mathcal{F}|\mathcal{G})}\sup_{\left(
a,y\right)  \in C\left(  x\right)  }\left\{  a+\mathbb{E}_{\mu}[yz|\mathcal{G}%
]+c\left(  z\right)  \right\}  .
\end{align*}
The third equality and the last inequality follow respectively from
\textit{i)} and \textit{v)} of Lemma \ref{jollyroger} in the Appendix. Now,
observe that%
\begin{align*}
I_{\operatorname*{niv}}\left(  x\right)   &  \leq\inf_{z\in L^{\infty
}(\mathcal{F}|\mathcal{G})}\sup_{\left(  a,y\right)  \in C\left(  x\right)
}\left\{  a+\mathbb{E}_{\mu}[yz|\mathcal{G}]+c\left(  z\right)  \right\} \\
&  \leq\inf_{z\in\Delta(\mathcal{F}|\mathcal{G})}\sup_{\left(  a,y\right)  \in
C\left(  x\right)  }\left\{  a+\mathbb{E}_{\mu}[yz|\mathcal{G}]+c\left(
z\right)  \right\} \\
&  \leq\inf_{z\in\Delta(\mathcal{F}|\mathcal{G})}\left\{  \sup_{a\in
L^{0}(\mathcal{G})}\left\{  a+\mathbb{E}_{\mu}[(x-a)z|\mathcal{G}]+c\left(
z\right)  \right\}  \right\} \\
&  =\inf_{z\in\Delta(\mathcal{F}|\mathcal{G})}\left\{  \sup_{a\in
L^{0}(\mathcal{G})}\left\{  \mathbb{E}_{\mu}[xz|\mathcal{G}]+c\left(
z\right)  \right\}  \right\} \\
&  =\inf_{z\in\Delta(\mathcal{F}|\mathcal{G})}\left\{  \mathbb{E}_{\mu
}[xz|\mathcal{G}]+c\left(  z\right)  \right\}  .
\end{align*}
The third inequality follows from the following observations: each $z\in
\Delta(\mathcal{F}|\mathcal{G})$ takes only nonnegative values$;$ $(a,x-a)\in
C\left(  x\right)  $ for each $a\in L^{0}(\mathcal{G})$; $\mathbb{E}_{\mu
}[\cdot|\mathcal{G}]$ is monotone; $x-a\geq y$ for each $(a,y)\in C(x)$; and
from \textit{iii)} of Lemma \ref{jollyroger} in the Appendix. Finally, let
$\bar{z}\in\operatorname*{dom}c\cap\Delta(\mathcal{F}|\mathcal{G})$. Then, by
the monotonicity of the conditional expectation operator,%
\begin{align*}
I_{\operatorname*{niv}}\left(  x\right)   &  \leq\inf_{z\in\Delta
(\mathcal{F}|\mathcal{G})}\left\{  \mathbb{E}_{\mu}[xz|\mathcal{G}]+c\left(
z\right)  \right\} \\
&  \leq\mathbb{E}_{\mu}[x\bar{z}|\mathcal{G}]+c\left(  \bar{z}\right) \\
&  \leq\mathbb{E}_{\mu}[|x\bar{z}||\mathcal{G}]+c\left(  \bar{z}\right) \\
&  \leq\left\Vert x\right\Vert _{\infty}^{\mathcal{G}}\mathbb{E}_{\mu}[\bar
{z}|\mathcal{G}]+c\left(  \bar{z}\right) \\
&  =\left\Vert x\right\Vert _{\infty}^{\mathcal{G}}+c\left(  \bar{z}\right)
\in L^{0}(\mathcal{G}).
\end{align*}
For all $x\in L^{\infty}(\mathcal{F}|\mathcal{G})$, $(\mathbf{0}_{S},x)\in
C(x)$ and hence, by definition of $I_{\operatorname*{niv}},$%
\[
-\infty<I\left(  x\right)  \leq I_{\operatorname*{niv}}\left(  x\right)  .
\]
Thus $I_{\operatorname*{niv}}$ is $L^{0}(\mathcal{G})$-valued.\hfill
$\blacksquare$

\noindent\textbf{Proof of Lemma \ref{lem:con-niv}} By Proposition
\ref{pro:dua-niv} and the last step of the previous proof
$I_{\operatorname*{niv}}$ is $L^{0}(\mathcal{G})$-valued and
$I_{\operatorname*{niv}}\geq I$. Passing to $L^{0}(\mathcal{G})$-translation
invariance, notice that $(b,y)\in C(x+a)$ for some $a\in L^{0}(\mathcal{G})$
and $x\in L^{\infty}(\mathcal{F}|\mathcal{G})$ if and only if $(b-a,y)\in
L^{0}(\mathcal{G})\times L^{\infty}(\mathcal{F}|\mathcal{G})$ and $y+(b-a)\leq
x,$ that is, if and only if $(b-a,y)\in C(x).$ Thus, if $a\in L^{0}%
(\mathcal{G})$ and $x\in L^{\infty}(\mathcal{F}|\mathcal{G})$,\ then%
\begin{align*}
I_{\operatorname*{niv}}\left(  x+a\right)   &  =\sup_{\left(  b,y\right)  \in
C\left(  x+a\right)  }\left\{  b+I\left(  y\right)  \right\}  =\sup_{\left(
b-a,y\right)  \in C\left(  x\right)  }\left\{  b+I\left(  y\right)  \right\}
\\
&  =\sup_{\left(  b-a,y\right)  \in C\left(  x\right)  }\left\{
(b-a)+a+I\left(  y\right)  \right\} \\
&  =\sup_{\left(  b-a,y\right)  \in C\left(  x\right)  }\left\{
(b-a)+I\left(  y\right)  \right\}  +a\\
&  =I_{\operatorname*{niv}}\left(  x\right)  +a
\end{align*}
where the second-to-last equality follows from \textit{i)} of Lemma
\ref{jollyroger} in the Appendix. This yields that $I_{\operatorname*{niv}}$
is $L^{0}(\mathcal{G})$-translation invariant. Now, let $x_{1},x_{2}\in
L^{\infty}(\mathcal{F}|\mathcal{G})$. If $x_{1}\geq x_{2}$, we have that
$C\left(  x_{1}\right)  \supseteq C\left(  x_{2}\right)  $, yielding that
$I_{\operatorname*{niv}}\left(  x_{1}\right)  \geq I_{\operatorname*{niv}%
}\left(  x_{2}\right)  $ so $I_{\operatorname*{niv}}$\ is monotone. Therefore,
$I_{\operatorname*{niv}}$\ is an $L^{0}(\mathcal{G})$-niveloid.

To conclude we prove $L^{0}(\mathcal{G})$-concavity. Consider $a\in
L^{0}(\mathcal{G})$ with\textbf{ }$\mathbf{0}_{S}\mathbf{\leq}a\leq
\mathbf{1}_{S}$. Note that if $\left(  b_{1},y_{1}\right)  \in C\left(
x_{1}\right)  $ and $\left(  b_{2},y_{2}\right)  \in C\left(  x_{2}\right)  $,
then $\left(  ab_{1}+\left(  \mathbf{1}_{S}-a\right)  b_{2},ay_{1}+\left(
\mathbf{1}_{S}-a\right)  y_{2}\right)  \in C\left(  ax_{1}+\left(
\mathbf{1}_{S}-a\right)  x_{2}\right)  $. Then, we have%
\begin{align*}
I_{\operatorname*{niv}}\left(  ax_{1}+\left(  \mathbf{1}_{S}-a\right)
x_{2}\right)   &  =\sup_{\left(  b,y\right)  \in C\left(  ax_{1}+\left(
\mathbf{1}_{S}-a\right)  x_{2}\right)  }\left\{  b+I\left(  y\right)  \right\}
\\
&  \geq\sup_{\left(  b_{1},y_{1}\right)  \in C\left(  x_{1}\right)  ,\text{
}\left(  b_{2},y_{2}\right)  \in C\left(  x_{2}\right)  }\left\{
ab_{1}+\left(  \mathbf{1}_{S}-a\right)  b_{2}+I\left(  ay_{1}+\left(
\mathbf{1}_{S}-a\right)  y_{2}\right)  \right\} \\
&  \geq\sup_{\left(  b_{1},y_{1}\right)  \in C\left(  x_{1}\right)  ,\text{
}\left(  b_{2},y_{2}\right)  \in C\left(  x_{2}\right)  }\left\{  a\left(
b_{1}+I\left(  y_{1}\right)  \right)  +\left(  \mathbf{1}_{S}-a\right)
\left(  b_{2}+I\left(  y_{2}\right)  \right)  \right\} \\
&  \geq\sup_{\left(  b_{1},y_{1}\right)  \in C\left(  x_{1}\right)  }\left\{
a\left(  b_{1}+I\left(  y_{1}\right)  \right)  \right\}  +\sup_{\left(
b_{2},y_{2}\right)  \in C\left(  x_{2}\right)  }\left\{  \left(
\mathbf{1}_{S}-a\right)  \left(  b_{2}+I\left(  y_{2}\right)  \right)
\right\} \\
&  =a\sup_{\left(  b_{1},y_{1}\right)  \in C\left(  x_{1}\right)  }\left\{
b_{1}+I\left(  y_{1}\right)  \right\}  +\left(  \mathbf{1}_{S}-a\right)
\sup_{\left(  b_{2},y_{2}\right)  \in C\left(  x_{2}\right)  }\left\{
b_{2}+I\left(  y_{2}\right)  \right\} \\
&  =aI_{\operatorname*{niv}}\left(  x_{1}\right)  +\left(  \mathbf{1}%
_{S}-a\right)  I_{\operatorname*{niv}}\left(  x_{2}\right)
\end{align*}
where the second inequality follows from the discussion above, the third
inequality from the $L^{0}(\mathcal{G})$-concavity of $I$ and point
\textit{iii)} of Lemma \ref{jollyroger} in the Appendix, the fourth inequality
from point \textit{ii)} of Lemma \ref{jollyroger} in the Appendix, and the
fifth inequality from Lemma \ref{scaloutfromsup} in the Appendix. This proves
$L^{0}(\mathcal{G})$-concavity of $I_{\operatorname*{niv}}.$\hfill
$\blacksquare$

Given $I:L^{\infty}(\mathcal{F}|\mathcal{G})\rightarrow\bar{L}^{0}%
(\mathcal{G})$ as in Lemma \ref{lem:niv} we denote by $\mathcal{J}$ the family
of $L^{0}(\mathcal{G})$-niveloids that dominate pointwise $I$.

\noindent\textbf{Proof of Lemma \ref{lem:niv} }By Lemma \ref{lem:con-niv} we
have that $I_{\operatorname*{niv}}$ is an $L^{0}(\mathcal{G})$-niveloid such
that $I_{\operatorname*{niv}}\geq I$, yielding that $\mathcal{J\not =%
}\emptyset$. Define $\bar{J}:L^{\infty}(\mathcal{F}|\mathcal{G})\rightarrow
\bar{L}^{0}(\mathcal{G})$\ to be such that $\bar{J}\left(  x\right)
=\inf_{J\in\mathcal{J}}J\left(  x\right)  $ for all $x\in L^{\infty
}(\mathcal{F}|\mathcal{G})$.\ It is immediate to see that $\bar{J}$ is an
$L^{0}(\mathcal{G})$-niveloid such that $I_{\operatorname*{niv}}\geq\bar
{J}\geq I$. Next, fix $x\in L^{\infty}(\mathcal{F}|\mathcal{G})$. We have that
for all $J\in\mathcal{J}$%
\[
a+I\left(  y\right)  \leq a+J\left(  y\right)  \leq a+J\left(  x-a\right)
=J\left(  x\right)  \quad\forall\left(  a,y\right)  \in C\left(  x\right)  .
\]
By (\ref{eq:def-I-niv}) and since $x$ and $J$ were arbitrarily chosen, we can
conclude that%
\[
I_{\operatorname*{niv}}\left(  x\right)  =\sup_{\left(  a,y\right)  \in
C\left(  x\right)  }\left\{  a+I\left(  y\right)  \right\}  \leq J\left(
x\right)  \quad\forall x\in L^{\infty}(\mathcal{F}|\mathcal{G}),\forall
J\in\mathcal{J}%
\]
yielding that $I_{\operatorname*{niv}}\leq\bar{J}$ and, in particular,
$I_{\operatorname*{niv}}=\bar{J}$. Define $\tilde{J}:L^{\infty}(\mathcal{F}%
|\mathcal{G})\rightarrow L^{0}(\mathcal{G})$\ by $\tilde{J}\left(  x\right)
=\inf_{y\in\Delta(\mathcal{F}|\mathcal{G})}\left\{  \mathbb{E}_{\mu
}[xy|\mathcal{G}]+c\left(  y\right)  \right\}  $ for all $x\in L^{\infty
}(\mathcal{F}|\mathcal{G})$.\footnote{At the cost of being pedantic, notice
that this operator is well defined since we assumed that $\operatorname*{dom}%
c\cap\Delta(\mathcal{F}|\mathcal{G})\not =\emptyset.$ Indeed, by Proposition
2, $\tilde{J}\geq I_{\operatorname*{niv}}$, and%
\[
\tilde{J}\left(  x\right)  \leq\mathbb{E}_{\mu}[xz|\mathcal{G}]+c\left(
z\right)  \in L^{0}(\mathcal{G})
\]
\par
for all $z\in\operatorname*{dom}c\cap\Delta(\mathcal{F}|\mathcal{G})$ and all
$x\in L^{\infty}(\mathcal{F}|\mathcal{G}).$} By Proposition \ref{pro:dua-niv},
we have that\ $\tilde{J}\geq I_{\operatorname*{niv}}$. Define $d:L^{1}%
(\mathcal{F}|\mathcal{G})\rightarrow\bar{L}^{0}(\mathcal{G})$ by%
\[
d\left(  y\right)  =\sup_{z\in L^{\infty}(\mathcal{F}|\mathcal{G})}\left\{
I_{\operatorname*{niv}}\left(  z\right)  -\mathbb{E}_{\mu}[yz|\mathcal{G}%
]\right\}  \qquad\forall y\in L^{1}(\mathcal{F}|\mathcal{G}).
\]
Since $I_{\operatorname*{niv}}\geq I$,\ we have that for all $y\in
\Delta(\mathcal{F}|\mathcal{G})$%
\[
c\left(  y\right)  =\sup_{z\in L^{\infty}(\mathcal{F}|\mathcal{G})}\left\{
I\left(  z\right)  -\mathbb{E}_{\mu}[yz|\mathcal{G}]\right\}  \leq\sup_{z\in
L^{\infty}(\mathcal{F}|\mathcal{G})}\left\{  I_{\operatorname*{niv}}\left(
z\right)  -\mathbb{E}_{\mu}[yz|\mathcal{G}]\right\}  =d\left(  y\right)  .
\]
Now notice that, by Lemma \ref{lem:con-niv} and Proposition \ref{pro:dua-niv},
$I_{\operatorname*{niv}}$ is $L^{0}(\mathcal{G})$-valued, monotone,
$L^{0}(\mathcal{G})$-translation invariant, and $L^{0}(\mathcal{G})$-concave.
Monotonicity and $L^{0}(\mathcal{G})$-translation invariance yield,
\[
I_{\operatorname*{niv}}(x)\leq I_{\operatorname*{niv}}(y)+\left\Vert
x-y\right\Vert _{\infty}^{\mathcal{G}}%
\]
for all $x,y\in L^{\infty}(\mathcal{F}|\mathcal{G}),$ and hence, rearranging
and interchanging $x$ and $y$, $\left\vert I_{\operatorname*{niv}%
}(x)-I_{\operatorname*{niv}}(y)\right\vert \leq\left\Vert x-y\right\Vert
_{\infty}^{\mathcal{G}}$ for all $x,y\in L^{\infty}(\mathcal{F}|\mathcal{G}).$
Thus, $I_{\operatorname*{niv}}$ is $\left\Vert \cdot\right\Vert _{\infty
}^{\mathcal{G}}$-continuous, $L^{0}(\mathcal{G})$-concave, monotone, and
$L^{0}(\mathcal{G})$-translation invariant. Now since $I_{\operatorname*{niv}%
}$ is $L^{0}(\mathcal{G})$-valued$,$ trivially $\mathtt{int}_{\left\Vert
\cdot\right\Vert _{\infty}^{\mathcal{G}}}(\operatorname*{dom}%
I_{\operatorname*{niv}})=L^{\infty}(\mathcal{F}|\mathcal{G})$ and by Lemma
\ref{represlemma} in the Appendix, we have that $I_{\operatorname*{niv}%
}(x)=\inf_{y\in\Delta(\mathcal{F}|\mathcal{G})}\left\{  \mathbb{E}_{\mu
}[xy|\mathcal{G}]+d(y)\right\}  $ for all $x\in L^{\infty}(\mathcal{F}%
|\mathcal{G})$. Consequently,%
\[
\tilde{J}\left(  x\right)  =\inf_{y\in\Delta(\mathcal{F}|\mathcal{G})}\left\{
\mathbb{E}_{\mu}[xy|\mathcal{G}]+c\left(  y\right)  \right\}  \leq\inf
_{y\in\Delta(\mathcal{F}|\mathcal{G})}\left\{  \mathbb{E}_{\mu}[xy|\mathcal{G}%
]+d\left(  y\right)  \right\}  =I_{\operatorname*{niv}}\left(  x\right)
\]
for all $x\in L^{\infty}(\mathcal{F}|\mathcal{G})$ proving that $\tilde
{J}=I_{\operatorname*{niv}}$ and (\ref{eq:dua-sma-niv}).\hfill$\blacksquare
$\bigskip

\noindent\textbf{Proof of Lemma \ref{measrep} }Fix $y\in\Delta(\mathcal{F}%
|\mathcal{G})$, we define $\nu:\mathcal{F\rightarrow}[0,1]$ as%
\[
\nu(A)=\int_{S}\mathbb{E}_{\mu}[\mathbf{1}_{A}y|\mathcal{G}]d\mu.
\]
for all $A\in\mathcal{F}.$ Notice that $\nu|_{\mathcal{G}}=\mu|_{\mathcal{G}}$
and that $\nu$ is a probability measure over $\mathcal{F}$.\footnote{Since
$y\in\Delta(\mathcal{F}|\mathcal{G})$ and $S\in\mathcal{G}$
\par%
\[
\nu(S)=\int_{S}\mathbb{E}_{\mu}[\mathbf{1}_{S}y|\mathcal{G}]d\mu=\int
_{S}\mathbb{E}_{\mu}[y|\mathcal{G}]d\mu=1.
\]
\par
and
\par%
\[
\nu\geq0.
\]
\par
Let $(A_{k})_{k=1}^{n}$be a finite collection of disjoint $\mathcal{F}%
$-measurable sets. Then%
\[
\nu(\cup_{k=1}^{n}A_{k})=\int_{S}\mathbb{E}_{\mu}[\mathbf{1}_{\cup_{k=1}%
^{n}A_{k}}y|\mathcal{G}]d\mu=\sum_{k=1}^{n}\int_{S}\mathbb{E}_{\mu}%
[\mathbf{1}_{A_{k}}y|\mathcal{G}]d\mu=\sum_{k=1}^{n}\nu(A_{k}).
\]
\par
Let $(A_{n})_{n\in%
\mathbb{N}
}$ be a countable collection of disjoint $\mathcal{F}$-measurable sets. Fix
$n\in%
\mathbb{N}
,$ $B_{n}=\cup_{k=1}^{n}A_{k},$ $B_{n}\uparrow\cup_{n\in%
\mathbb{N}
}A_{n}$ by both, the conditional and unconditional, versions of the Monotone
Convergence Theorem%
\[
\sum_{k=1}^{n}\nu(A_{k})=\nu(B_{n})=\int_{S}\mathbb{E}_{\mu}[\mathbf{1}%
_{B_{n}}y|\mathcal{G}]d\mu{\Huge \uparrow}\int_{S}\mathbb{E}_{\mu}%
[\mathbf{1}_{\cup_{n\in%
\mathbb{N}
}A_{n}}y|\mathcal{G}]d\mu=\nu(\cup_{n\in%
\mathbb{N}
}A_{n}).
\]
\par
Thus $\nu$ is a probability measure.
\par
{}} Let $B\in\mathcal{G},$ then%
\begin{align*}
\mathbb{E}_{\nu}[\mathbf{1}_{B}\mathbb{E}_{\mu}[\mathbf{1}_{A}y|\mathcal{G}]]
&  =\mathbb{E}_{\nu}[\mathbb{E}_{\mu}[\mathbf{1}_{A\cap B}y|\mathcal{G}%
]]=\int_{S}\mathbb{E}_{\mu}[\mathbf{1}_{A\cap B}y|\mathcal{G}]d\nu\\
&  =\int_{S}\mathbf{1}_{B}\mathbb{E}_{\mu}[\mathbf{1}_{A}y|\mathcal{G}%
]d\nu=\int_{B}\mathbb{E}_{\mu}[\mathbf{1}_{A}y|\mathcal{G}]d\nu\\
&  =\int_{B}\mathbb{E}_{\mu}[\mathbf{1}_{A}y|\mathcal{G}]d\mu=\int
_{S}\mathbb{E}_{\mu}[\mathbf{1}_{A\cap B}y|\mathcal{G}]d\mu\\
&  =\nu(A\cap B)=\mathbb{E}_{\nu}[\mathbf{1}_{A\cap B}]=\mathbb{E}_{\nu
}[\mathbf{1}_{B}\mathbb{E}_{\nu}[\mathbf{1}_{A}|\mathcal{G}]].
\end{align*}
Since $B\in\mathcal{G}$ was chosen arbitrarily, we have$\mathbb{\ \mathbb{E}%
_{\mu}[}\mathbf{1}_{A}y|\mathcal{G}]=\mathbb{E}_{\nu}\mathbb{[}\mathbf{1}%
_{A}|\mathcal{G}],$ indeed notice that both $\mathbb{\mathbb{E}_{\mu}%
[}\mathbf{1}_{A}y|\mathcal{G}],\mathbb{E}_{\nu}\mathbb{[}\mathbf{1}%
_{A}|\mathcal{G}]\geq0,$ and their integrals coincide over all $B\in
\mathcal{G}.$ Now, fix a simple random variable $x\in L^{0}(\mathcal{F}),$
then there exist a finite collection $(A_{k})_{k=1}^{n}$ of $\mathcal{F}%
$-measurable sets and a vector $(x_{k})_{k=1}^{n}\in\mathbb{R}^{n}$ such that%
\[
x=\sum_{k=1}^{n}x_{k}\mathbf{1}_{A_{k}}.
\]
Then,
\begin{equation}
\mathbb{\mathbb{E}_{\mu}[}xy|\mathcal{G}]=\mathbb{\mathbb{E}_{\mu}}\left[
\sum_{k=1}^{n}x_{k}\mathbf{1}_{A_{k}}y{\Huge |}\mathcal{G}\right]  =\sum
_{k=1}^{n}x_{k}\mathbb{\mathbb{E}_{\mu}[}\mathbf{1}_{A_{k}}y|\mathcal{G}%
]=\sum_{k=1}^{n}x_{k}\mathbb{E}_{\nu}\mathbb{[}\mathbf{1}_{A_{k}}%
|\mathcal{G}]=\mathbb{E}_{\nu}\mathbb{[}x|\mathcal{G}]. \label{eq:simple}%
\end{equation}
Now fix $x\in L^{\infty}(\mathcal{F}|\mathcal{G})_{+},$ then there exists a
sequence of simple random variables $(s_{n})_{n\in%
\mathbb{N}
}$ in $L^{0}(\mathcal{F})$ such that $s_{n}\uparrow x.$ Then, by Monotone
Convergence Theorem (for conditional expectations) and the equality
(\ref{eq:simple}) we have%
\[
\mathbb{\mathbb{E}_{\mu}[}xy|\mathcal{G}]=\lim_{n\rightarrow\infty
}\mathbb{\mathbb{E}_{\mu}[}s_{n}y|\mathcal{G}]=\lim_{n\rightarrow\infty
}\mathbb{\mathbb{E}_{\nu}[}s_{n}|\mathcal{G}]=\mathbb{\mathbb{E}_{\nu}%
[}x|\mathcal{G}].
\]
Repeating this argument for the positive and negative part of each $x\in
L^{\infty}(\mathcal{F}|\mathcal{G})$, we have%
\[
\mathbb{\mathbb{E}_{\mu}[}xy|\mathcal{G}]=\mathbb{\mathbb{E}_{\nu}%
[}x|\mathcal{G}]
\]
for all $x\in L^{\infty}(\mathcal{F}|\mathcal{G}).$\textbf{ }

Passing to the bijectivity part, notice first that $T$ is well defined, since
if $\nu_{1}=\nu_{2}$ with $\nu_{1},\nu_{2}\in\mathcal{M(G)},$ then there exist
$y_{\nu_{1}},y_{\nu_{2}}\in\Delta(\mathcal{F}|\mathcal{G})$
\[
\int_{A}y_{\nu_{1}}d\mu=\int_{A}y_{\nu_{2}}d\mu
\]
for all $A\in\mathcal{F}.$ Thus, $y_{\nu_{1}}=y_{\nu_{2}}.$ If $y\in
\Delta(\mathcal{F}|\mathcal{G}),$ we just proved that there exists, a $\nu
\in\mathcal{M(G)}$ represented by it, just let $\nu(A)=\int_{A}yd\mu$ for all
$A\in\mathcal{F}.$ Thus, $T$ is surjective. Moreover, let $T(\nu_{1}%
),T(\nu_{2})\in\Delta(\mathcal{F}|\mathcal{G})=T(\mathcal{M(G)}),$ if
$T(\nu_{1})=T(\nu_{2}),$ it follows that%
\[
\nu_{1}(A)=\int_{A}T(\nu_{1})d\mu=\int_{A}T(\nu_{2})d\mu=\nu_{2}(A)
\]
for all $A\in\mathcal{F}.$ Thus, $T$ is a bijection.\hfill$\blacksquare
$\bigskip

\noindent\textbf{Proof of Lemma \ref{charactM} }If $\nu\in\mathcal{M(G)}$,
then there exists $y\in\Delta(\mathcal{F}|\mathcal{G})$ such that $\nu
(A)=\int_{A}yd\mu.$ Since $\Delta(\mathcal{F}|\mathcal{G})\subseteq Y$, it
follows $\mathcal{M(G)\subseteq}\widehat{\mathcal{M}}\mathcal{(G)}.$
Conversely, suppose $\nu\in\widehat{\mathcal{M}}\mathcal{(G)},$ and denote by
$y_{\nu}$ the associated element of $Y.$ Then, for all $B\in\mathcal{G}$, it
follows that $\nu(B)=\mu(B)$ and%
\[
\int_{B}\mathbb{E}_{\mu}[y_{\nu}|\mathcal{G}]d\mu=\int_{B}y_{\nu}d\mu
=\nu(B)=\mu(B)=\int_{B}\mathbf{1}_{S}d\mu
\]
thus, since $\mathbb{E}_{\mu}[y_{\nu}|\mathcal{G}]$ and $\mathbf{1}_{S}$ are
both nonnegative and $\mathcal{G}$-measurable, it follows that $\mathbb{E}%
_{\mu}[y_{\nu}|\mathcal{G}]=\mathbf{1}_{S}$ and hence $y_{\nu}\in
\Delta(\mathcal{F}|\mathcal{G}).$ Therefore, $\widehat{\mathcal{M}%
}\mathcal{(G)=M(G)}.$\hfill$\blacksquare$\bigskip

\noindent\textbf{Proof of Proposition \ref{measrepniv} }Combining Lemma
\ref{measrep} with the representation result $I_{\operatorname*{niv}}%
(\cdot)=\inf_{y\in\Delta(\mathcal{F}|\mathcal{G})}\left\{  \mathbb{E}_{\mu
}[(\cdot)y|\mathcal{G}]+c\left(  y\right)  \right\}  $ proved in Lemma
\ref{lem:niv}, it follows that%
\[
I_{\operatorname*{niv}}\left(  x\right)  =\inf_{\nu\in\mathcal{M(G)}}\left\{
\mathbb{E}_{\nu}[x|\mathcal{G}]+\hat{c}\left(  \nu\right)  \right\}
\]
for all $x\in L^{\infty}(\mathcal{F}|\mathcal{G}).$\hfill$\blacksquare
$\bigskip

\noindent\textbf{Proof of Proposition \ref{orderofniv} }Fix $x\in L^{\infty
}(\mathcal{F}|\mathcal{G})$. If $\left(  a^{\prime},y^{\prime}\right)  \in
C\left(  x\right)  $. We have that%
\begin{align*}
a^{\prime}+I\left(  y^{\prime}\right)   &  \leq a^{\prime}+I\left(  \left(
y^{\prime}+a^{\prime}\right)  -a^{\prime}\right)  \leq\sup_{a\in
L^{0}(\mathcal{G})}\left\{  a+I\left(  \left(  y^{\prime}+a^{\prime}\right)
-a\right)  \right\} \\
&  \leq\sup_{\left\{  y\in L^{\infty}(\mathcal{F}|\mathcal{G}):y\leq
x\right\}  }\sup_{a\in L^{0}(\mathcal{G})}\left\{  a+I\left(  y-a\right)
\right\}
\end{align*}
yielding that
\[
I_{\operatorname*{niv}}\left(  x\right)  =\sup_{\left(  a,y\right)  \in
C\left(  x\right)  }\left\{  a+I\left(  y\right)  \right\}  \leq\sup_{\left\{
y\in L^{\infty}(\mathcal{F}|\mathcal{G}):y\leq x\right\}  }\sup_{a\in
L^{0}(\mathcal{G})}\left\{  a+I\left(  y-a\right)  \right\}  .
\]
Fix\ $L^{\infty}(\mathcal{F}|\mathcal{G})\ni y^{\prime}\leq x$ and $a^{\prime
}\in L^{0}(\mathcal{G})$. It follows that $a^{\prime}\in L^{0}(\mathcal{G}%
)$\ and $y^{\prime}-a^{\prime}\leq x-a^{\prime}$. We have that%
\[
a^{\prime}+I\left(  y^{\prime}-a^{\prime}\right)  \leq\sup_{\left(
a,y\right)  \in C\left(  x\right)  }\left\{  a+I\left(  y\right)  \right\}
\]
yielding that $\sup_{\left\{  y\in L^{\infty}(\mathcal{F}|\mathcal{G}):y\leq
x\right\}  }\sup_{a\in L^{0}(\mathcal{G})}\left\{  a+I\left(  y-a\right)
\right\}  \leq\sup_{\left(  a,y\right)  \in C\left(  x\right)  }\left\{
a+I\left(  y\right)  \right\}  =I_{\operatorname*{niv}}\left(  x\right)  $,
proving the statement. Notice that the lattice operations performed above are
legit since $I_{\operatorname*{niv}}$ maps into the complete lattice $\bar
{L}^{0}(\mathcal{G}).$\hfill$\blacksquare$\bigskip

\noindent\textbf{Proof of Corollary \ref{Rep_entropic_coce} }The proof follows
from first-order conditions and observing that the \textit{pointwise supremum
}is larger or equal than the \textit{lattice supremum}. First of all notice
that $\phi^{\ast}:t\mapsto e^{t}-1$. Thus, $a-\mathbb{E}_{\mu}[\phi^{\ast
}\left(  a-x\right)  |\mathcal{G}]=a-\mathbb{E}_{\mu}[e^{a-x}-1|\mathcal{G}%
]=a-e^{a}\mathbb{E}_{\mu}[e^{-x}|\mathcal{G}]+1$ for all $a\in L^{0}%
(\mathcal{G})$ and $x\in L^{\infty}(\mathcal{F}|\mathcal{G}).$ Fix $s\in S$
and $x\in L^{\infty}(\mathcal{F}|\mathcal{G}).$ Taking the first-order
condition with respect to $a(s)$ of $a(s)-e^{a(s)}\mathbb{E}_{\mu}%
[e^{-x}|\mathcal{G}](s)+1$ which yields,%
\[
a(s)=-\log\left(  \mathbb{E}_{\mu}[e^{-x}|\mathcal{G}](s)\right)  .
\]
Thanks to this expression we have that%
\[
\sup_{a\in L^{0}(\mathcal{G})}\left\{  a(s)-e^{a(s)}\mathbb{E}_{\mu}%
[e^{-x}|\mathcal{G}](s)+1\right\}  =-\log\left(  \mathbb{E}_{\mu}%
[e^{-x}|\mathcal{G}](s)\right)  .
\]
Now notice that $s\mapsto-\log\left(  \mathbb{E}_{\mu}[e^{-x}|\mathcal{G}%
](s)\right)  $ is $\mathcal{G}$-measurable and since $x\in L^{\infty
}(\mathcal{F}|\mathcal{G})$, we have $-\log\left(  \mathbb{E}_{\mu}%
[e^{-x}|\mathcal{G}]\right)  \in L^{0}(\mathcal{G})$. Given that $\sup_{a\in
L^{0}(\mathcal{G})}\left\{  a(s)-e^{a(s)}\mathbb{E}_{\mu}[e^{-x}%
|\mathcal{G}](s)+1\right\}  \geq\sup_{a\in L^{0}(\mathcal{G})}\left\{
a-e^{a}\mathbb{E}_{\mu}[e^{-x}|\mathcal{G}]+1\right\}  (s)$ for all $s\in S$,
we have that%
\[
\sup_{a\in L^{0}(\mathcal{G})}\left\{  a-\mathbb{E}_{\mu}[\phi^{\ast}\left(
a-x\right)  |\mathcal{G}]\right\}  =\sup_{a\in L^{0}(\mathcal{G})}\left\{
a-e^{a}\mathbb{E}_{\mu}[e^{-x}|\mathcal{G}]+1\right\}  =-\log\left(
\mathbb{E}_{\mu}[e^{-x}|\mathcal{G}]\right)  .
\]
By Theorem \ref{Rep_COCEs} the claim follows.\hfill$\blacksquare$

\subsection{A toolbox}

In this subsection of the Appendix, we report some simple results that we used
frequently in the proofs above and in the rest of the Appendix. While these
results are very simple, reporting them explicitly, with their proofs, makes
the exposition more transparent.

\begin{lemma}
\label{jollyroger}Let $(X,\leq)$ be a Riesz space, then we have that,

i) for all nonempty $Y\subseteq X$ such that $\sup Y$ exists, we have
\[
\sup\left\{  x+Y\right\}  =x+\sup Y
\]
for all $x\in X.$

ii) Let $A$, $B$ be nonempty sets and consider two functions $f:A\rightarrow
X,$ $g:B\rightarrow X$ such that $\sup_{a\in A}f(a),$ $\sup_{b\in B}g(b),$ and
$\sup_{(a,b)\in A\times B}\{f(a)+g(b)\}$ exist. Then,%
\[
\sup_{a\in A}f(a)+\sup_{b\in B}g(b)\leq\sup_{(a,b)\in A\times B}%
\{f(a)+g(b)\}.
\]

iii) Let $A$ be a nonempty set and $f,g:A\rightarrow X$ with $f(a)\leq g(a)$
for all $a\in A.$ If $\sup_{a\in A}f(a)$ and $\sup_{a\in A}g(a)$ exist, then,%
\[
\sup_{a\in A}f(a)\leq\sup_{a\in A}g(a).
\]

iv) Let $A,$ $B$ be nonempty sets and $f,g:A\times B\rightarrow X$ with
$f(a,b)\leq g(a,b)$ for all $a\in A$ and all $b\in B.$ If $\inf_{b\in B}%
\sup_{a\in A}f(a,b)$ and $\inf_{b\in B}\sup_{a\in A}g(a,b)$ exist, then,%
\[
\inf_{b\in B}\sup_{a\in A}f(a,b)\leq\inf_{b\in B}\sup_{a\in A}g(a,b).
\]

v) Let $A$, $B$ be nonempty sets and consider a function $f:A\times
B\rightarrow X$, such that $\sup_{a\in A}f(a,b)$, $\inf_{b\in B}f(a,b),$
$\sup_{a\in A}\inf_{b\in B}f(a,b),$ and $\inf_{b\in B}\sup_{a\in A}f(a,b)$
exist for all $a\in A$ and $b\in B$. Then,%
\[
\sup_{a\in A}\inf_{b\in B}f(a,b)\leq\inf_{b\in B}\sup_{a\in A}f(a,b).
\]

\end{lemma}

\noindent\textbf{Proof }\textit{i)} Since $X$ is a Riesz space $x\leq y$ if
and only if $x+z\leq y+z$ for all $z\in X.$ This yields that\ $x+y\leq x+\sup
Y$ for all $x$,$y\in Y,$ and hence $\sup\left\{  x+Y\right\}  \leq x+\sup Y.$
To conclude, $x+y\leq\sup\left\{  x+Y\right\}  $\ for all $x$,$y\in Y$\ and
hence $x+\sup Y\leq\sup\left\{  x+Y\right\}  .$

\textit{ii)} Fix $a^{\prime}\in A,$ $b^{\prime}\in B.$ Then, $f(a^{\prime
})+g(b^{\prime})\leq\sup_{(a,b)\in A\times B}\{f(a)+g(b)\}.$ Since,
$b^{\prime}$ was chosen arbitrarily, this yields $f(a^{\prime})+\sup_{b\in
B}g(b)\leq\sup_{(a,b)\in A\times B}\{f(a)+g(b)\}.$ Analogously, since
$a^{\prime}$ was chosen arbitrarily, we have $\sup_{a\in A}\{f(a)+\sup_{b\in
B}g(b)\}\leq\sup_{(a,b)\in A\times B}\{f(a)+g(b)\}.$ In conclusion, point
\textit{i)} yields the claim
\[
\sup_{a\in A}f(a)+\sup_{b\in B}g(b)\leq\sup_{(a,b)\in A\times B}%
\{f(a)+g(b)\}.
\]

\textit{iii)} For all $a^{\prime}\in A,$ $f(a^{\prime})\leq\sup_{a\in A}g(a),$
and hence $\sup_{a\in A}f(a)\leq\sup_{a\in A}g(a).$

\textit{iv)} We have that for all $a^{\prime}\in A$ and all $b\in B,$
$f(a^{\prime},b)\leq\sup_{a\in A}g(a,b).$ By point \textit{iii)}, $\inf_{b\in
B}\sup_{a\in A}$ $f(a,b)\leq\inf_{b\in B}\sup_{a\in A}g(a,b).$

\textit{v)} Fix $a^{\prime}\in A,$ $b^{\prime}\in B.$ Then, $f(a^{\prime
},b^{\prime})\leq\sup_{a\in A}f(a,b^{\prime}).$ Clearly $\inf_{b\in B}$
$f(a^{\prime},b)\leq\sup_{a\in A}f(a,b^{\prime})$ and since $a^{\prime}$ was
chosen arbitrarily,
\[
\sup_{a\in A}\inf_{b\in B}f(a,b)\leq\sup_{a\in A}f(a,b^{\prime}).
\]
Analogously, since $b^{\prime}$ was chosen arbitrarily, we have%
\[
\sup_{a\in A}\inf_{b\in B}f(a,b)\leq\inf_{b\in B}\sup_{a\in A}f(a,b).
\]
\hfill\hfill$\blacksquare$

We have also the following.\footnote{In what follows, we adopt the notation
$\left\{  a\gtreqless r\right\}  $ to denote $\left\{  s\in S:a(s)\gtreqless
r\right\}  $ for all $a\in L^{0}(\mathcal{G})$ and all $r\in\mathbb{R}$.}

\begin{lemma}
\label{scaloutfromsup}Let $L\subseteq L^{0}(\mathcal{G})$ be such that $\sup
L$ exists in $L^{0}(\mathcal{G})$, then%
\[
\forall a\in L^{0}(\mathcal{G})_{+},\text{ }\sup aL=a\sup L.
\]

\end{lemma}

\noindent\textbf{Proof }Let $a\in L^{0}(\mathcal{G})_{+}.$ Then, for all $x\in
L,$ we have $\sup L\geq x$ and hence $a\sup L\geq\sup aL$. Therefore, $a\sup
L$ is an upper bound of $aL.$ Let $b\in L^{0}(\mathcal{G})$ be an upper bound
of $aL,$ then $b\geq ac,$ for some $c\in L^{0}(\mathcal{G})$ with $c\geq x$
for all $x\in L.$\footnote{Indeed, notice that if $a=\mathbf{0}_{S},$ then the
claim is trivial. Therefore, assume that $a\neq\mathbf{0}_{S}.$ It follows
that $\mu(\{a>0\})>0.$ By hypothesis $b\geq ax$ for all $x\in L.$ Thus,
$b|_{\{a>0\}}\geq ax|_{\{a>0\}}=a|_{\{a>0\}}x|_{\{a>0\}}.$ This yields that
$\frac{b|_{\{a>0\}}}{a|_{\{a>0\}}}\geq x|_{\{a>0\}}$ for all $x\in L.$ Then,
define%
\[
c(s)=\left\{
\begin{array}
[c]{cc}%
\frac{b|_{\{a>0\}}}{a|_{\{a>0\}}}(s) & \text{\emph{if }}s\in\{a>0\}\\
(\sup L)(s) & \text{\emph{otherwise}}%
\end{array}
\right.
\]
\par
for all $s\in S.$ Clearly, $c\in$ $L^{0}(\mathcal{G})$ and $c\geq x$ for all
$x\in L.$ In addition, if $s\in\{a>0\},$ then $b(s)=a(s)c(s).$ On the other
hand if $s\in\{a=0\}$, since $b$ is an upper bound of $aL,$ for all $x\in L,$
$b(s)\geq a(s)x(s)=0=$ $a(s)c(s).$ Thus, $b\geq ac.$} Then consider
$d=\mathbf{1}_{\{a>0\}}c+\mathbf{1}_{\{a=0\}}\sup L.$ Clearly, we have that
$d\geq x$ for all $x\in L$ and hence $ad\geq a\sup L.$ Therefore, we have
$b\geq ac=ad\geq a\sup L.$ Therefore, $a\sup L$ is the smallest upper bound of
$aL,$ that is $a\sup L=\sup aL.$\hfill$\blacksquare$

In turn this lemma yields also the following%
\[
\mathbf{1}_{A}\sup L=\sup\mathbf{1}_{A}L=\sup\mathbf{1}_{A}\mathbf{1}%
_{A}L=\mathbf{1}_{A}\sup\mathbf{1}_{A}L
\]
for all $L\subseteq L^{0}(\mathcal{G})$ such that $\sup L$ exists in
$L^{0}(\mathcal{G})$ and all $A\in\mathcal{G}.$

\subsection{Representation results for proper functions}

Here, we report some concave duality results. These duality representations
are adapted from more general results in \cite{AppCondRisk} and \cite{Guo}. We
report them with their related proofs for the sake of completeness and self-containment.

\begin{definition}
A proper function $f:L^{\infty}(\mathcal{F}|\mathcal{G})\rightarrow\bar{L}%
^{0}(\mathcal{G})$ is said to be local if $\mathbf{1}_{A}f(\mathbf{1}%
_{A}x)=\mathbf{1}_{A}f(x)$ for all $x\in L^{\infty}(\mathcal{F}|\mathcal{G})$
and all $A\in\mathcal{G}.$
\end{definition}

We report, with the related proof, a standard characterization of local property.

\begin{lemma}
\label{regularity} Let $f:L^{\infty}(\mathcal{F}|\mathcal{G})\rightarrow
\bar{L}^{0}(\mathcal{G})$ be a proper function$.$ Then, $f$ is local if and
only if it is regular, i.e.,%
\[
f(\mathbf{1}_{A}x+\mathbf{1}_{A^{c}}y)=\mathbf{1}_{A}f(x)+\mathbf{1}_{A^{c}%
}f(y)
\]
for all $x,y\in L^{\infty}(\mathcal{F}|\mathcal{G})$ and all $A\in
\mathcal{G}.$
\end{lemma}

\noindent\textbf{Proof }Suppose $f$ is local, $x,y\in L^{\infty}%
(\mathcal{F}|\mathcal{G}),$ and $A\in\mathcal{G}.$ First of all notice that%
\[
f(\mathbf{1}_{A}x)=f(\mathbf{1}_{A}\left(  \mathbf{1}_{A}x+\mathbf{1}_{A^{c}%
}y\right)  )\text{ and }f(\mathbf{1}_{A^{c}}y)=f(\mathbf{1}_{A^{c}}\left(
\mathbf{1}_{A}x+\mathbf{1}_{A^{c}}y\right)  ).
\]
Then, by the localness of $f$ we have%
\[
\mathbf{1}_{A}f(\mathbf{1}_{A}\left(  \mathbf{1}_{A}x+\mathbf{1}_{A^{c}%
}y\right)  )=\mathbf{1}_{A}f(\mathbf{1}_{A}x+\mathbf{1}_{A^{c}}y)
\]
and
\[
\mathbf{1}_{A^{c}}f(\mathbf{1}_{A^{c}}\left(  \mathbf{1}_{A}x+\mathbf{1}%
_{A^{c}}y\right)  )=\mathbf{1}_{A^{c}}f(\mathbf{1}_{A}x+\mathbf{1}_{A^{c}}y).
\]
Combining the equalities proved above and exploiting the localness of $f$, we
have%
\begin{align*}
\mathbf{1}_{A}f(x)+\mathbf{1}_{A^{c}}f(y)  &  =\mathbf{1}_{A}f(\mathbf{1}%
_{A}x)+\mathbf{1}_{A^{c}}f(\mathbf{1}_{A^{c}}y)\\
&  =\mathbf{1}_{A}f(\mathbf{1}_{A}\left(  \mathbf{1}_{A}x+\mathbf{1}_{A^{c}%
}y\right)  )+\mathbf{1}_{A^{c}}f(\mathbf{1}_{A^{c}}\left(  \mathbf{1}%
_{A}x+\mathbf{1}_{A^{c}}y\right)  )\\
&  =\mathbf{1}_{A}f(\mathbf{1}_{A}x+\mathbf{1}_{A^{c}}y)+\mathbf{1}_{A^{c}%
}f(\mathbf{1}_{A}x+\mathbf{1}_{A^{c}}y)\\
&  =f(\mathbf{1}_{A}x+\mathbf{1}_{A^{c}}y).
\end{align*}
Thus, $f$ is regular.

To prove the converse suppose that $f$ is regular, $x\in L^{\infty
}(\mathcal{F}|\mathcal{G}),$ and $A\in\mathcal{G}.$ Then,%
\[
f(\mathbf{1}_{A}x)=f(\mathbf{1}_{A}x+\mathbf{1}_{A^{c}}\mathbf{0}%
_{S})=\mathbf{1}_{A}f(x)+\mathbf{1}_{A^{c}}f(\mathbf{0}_{S})
\]
multiplying both sides by $\mathbf{1}_{A}$ the result follows.\hfill
$\blacksquare$

Moreover, we have that $L^{0}(\mathcal{G})$-concavity implies the localness of
the function. The next lemma is due to \cite{SepandDual} (see Theorem 3.2).

\begin{lemma}
\label{concimplloc} Let $f:L^{\infty}(\mathcal{F}|\mathcal{G})\rightarrow
\bar{L}^{0}(\mathcal{G})$ be proper and $L^{0}(\mathcal{G})$-concave. Then,
$f$ is local.
\end{lemma}

\noindent\textbf{Proof }Let $A\in\mathcal{G}$ and $x\in L^{\infty}%
(\mathcal{F}|\mathcal{G})$. Then, we have that%
\begin{align*}
f(\mathbf{1}_{A}x)  &  =f(\mathbf{1}_{A}x+\mathbf{1}_{A^{c}}\mathbf{0}_{S})\\
&  \geq\mathbf{1}_{A}f(x)+\mathbf{1}_{A^{c}}f(\mathbf{0}_{S})\\
&  =\mathbf{1}_{A}f(\mathbf{1}_{A}(\mathbf{1}_{A}x)+\mathbf{1}_{A^{c}%
}x)+\mathbf{1}_{A^{c}}f(\mathbf{0}_{S})\\
&  \geq\mathbf{1}_{A}f(\mathbf{1}_{A}x)+\mathbf{1}_{A^{c}}f(\mathbf{0}_{S})
\end{align*}

multiplying both sides by $\mathbf{1}_{A}$ the result follows.\hfill
$\blacksquare$

Notice that Lemma \ref{concimplloc} holds even if $f$ is asked to be
$L^{0}(\mathcal{G})$-convex. The following result is due to \cite{AppCondRisk}
(see the discussion after Definition 3.5). First of all, notice that for all
sequences $(x_{n})_{n\in\mathbb{N}}$ in $L^{\infty}(\mathcal{F}|\mathcal{G})$
and all countable partitions of $S$ of $\mathcal{G}$-measurable sets
$(A_{n})_{n\in\mathbb{N}}$, there exists $x$ in $L^{\infty}(\mathcal{F}%
|\mathcal{G})$ such that $\mathbf{1}_{A_{n}}x_{n}=\mathbf{1}_{A_{n}}x$ for all
$n\in\mathbb{N}$.\footnote{Take $x=\sum_{n\in\mathbb{N}}\mathbf{1}_{A_{n}%
}x_{n}$. Then,%
\[
\mathbf{1}_{A_{k}}x=\sum_{n\in\mathbb{N}}\mathbf{1}_{A_{k}}\mathbf{1}_{A_{n}%
}x_{n}\mathbf{=1}_{A_{k}}x_{k}%
\]
\par
for all $k\in\mathbb{N}$ and,
\par%
\[
|x|\leq\sum_{n\in\mathbb{N}}\mathbf{1}_{A_{n}}\left\vert x_{n}\right\vert
\leq\sum_{n\in\mathbb{N}}\mathbf{1}_{A_{n}}\left\Vert x_{n}\right\Vert
_{\infty}^{\mathcal{G}}\in L^{0}(\mathcal{G})
\]
\par
where the last implication follows from the fact $\left\Vert x_{n}\right\Vert
_{\infty}^{\mathcal{G}}\in L^{0}(\mathcal{G})$ for all $n\in\mathbb{N}$ and
$(A_{n})_{n\in\mathbb{N}}$ is a partition of $S$ of $\mathcal{G}$-measurable
sets.} In particular, we say that $L^{\infty}(\mathcal{F}|\mathcal{G})$
satisfies the \textit{countable concatenation property. }For a function
$f:L^{\infty}(\mathcal{F}|\mathcal{G})\rightarrow\bar{L}^{0}(\mathcal{G})$ we
define the map $c_{f}$ as $c_{f}(y)=\sup_{w\in L^{\infty}(\mathcal{F}%
|\mathcal{G})}\left\{  f(w)-\mathbb{E}_{\mu}[wy|\mathcal{G}]\right\}  $ for
all $y\in L^{1}(\mathcal{F}|\mathcal{G})$ while by $f^{\ast}$ we denote the
classic convex conjugate, $f^{\ast}(y)=\sup_{w\in L^{\infty}(\mathcal{F}%
|\mathcal{G})}\left\{  \mathbb{E}_{\mu}[wy|\mathcal{G}]-f\left(  w\right)
\right\}  $ for all $y\in L^{1}(\mathcal{F}|\mathcal{G}).$

\begin{lemma}
\label{domina proper}Let $f:L^{\infty}(\mathcal{F}|\mathcal{G})\rightarrow
\bar{L}^{0}(\mathcal{G})$ be a proper, $L^{0}(\mathcal{G})$-concave,
$\left\Vert \cdot\right\Vert _{\infty}^{\mathcal{G}}$-continuous function with
$x_{0}\in\operatorname*{dom}f$. Then,%
\[
c_{f}(y)=\sup_{w\in\operatorname*{dom}f}\left\{  f(w)-\mathbb{E}_{\mu
}[wy|\mathcal{G}]\right\}
\]
and, if $x_{0}\in\mathtt{int}_{\left\Vert \cdot\right\Vert _{\infty
}^{\mathcal{G}}}(\operatorname*{dom}f),$
\[
f(x)=\inf_{y\in\operatorname*{dom}c_{f}}\left\{  \mathbb{E}_{\mu
}[xy|\mathcal{G}]+c_{f}\left(  y\right)  \right\}
\]
for all $y\in L^{1}(\mathcal{F}|\mathcal{G})$ and all $x\in L^{\infty
}(\mathcal{F}|\mathcal{G}).$
\end{lemma}

\noindent\textbf{Proof }Let $y\in L^{1}(\mathcal{F}|\mathcal{G}).$ If
$x\in\operatorname*{dom}f,$ then,
\[
f\left(  x\right)  -\mathbb{E}_{\mu}[xy|\mathcal{G}]\leq\sup_{w\in
\operatorname*{dom}f}\left\{  f(w)-\mathbb{E}_{\mu}[wy|\mathcal{G}]\right\}
.
\]
If $x\notin\operatorname*{dom}f,$ then, since $f<\infty,$ it is enough to
consider $A_{x}=\{f(x)=-\infty\}.$ In particular, $A_{x}$ is a $\mathcal{G}%
$-measurable set with $\mu(A_{x})>0.$ Then, we have%
\begin{align*}
\mathbf{1}_{A_{x}}[f(x)-\mathbb{E}_{\mu}[xy|\mathcal{G}]]  &  =\mathbf{1}%
_{A_{x}}\cdot(-\infty)\\
&  \leq\mathbf{1}_{A_{x}}\sup_{w\in\operatorname*{dom}f}\left\{
f(w)-\mathbb{E}_{\mu}[wy|\mathcal{G}]\right\}  \text{.}%
\end{align*}
Since $f$ is $L^{0}(\mathcal{G})$-concave, then by Lemma \ref{concimplloc} it
is local and by Lemma \ref{regularity} it is also regular. Then, we have that%
\[
f(\mathbf{1}_{A_{x}^{c}}x+\mathbf{1}_{A_{x}}x_{0})=\mathbf{1}_{A_{x}^{c}%
}f(x)+\mathbf{1}_{A_{x}}f(x_{0}).
\]
Since $x_{0}\in\operatorname*{dom}f$, $\mathbf{1}_{A_{x}^{c}}f(x)>-\infty,$
and $f$ is proper, it follows
\[
\mathbf{1}_{A_{x}^{c}}x+\mathbf{1}_{A_{x}}x_{0}\in\operatorname*{dom}f.
\]
This yields $\mathbf{1}_{A_{x}^{c}}x=\mathbf{1}_{A_{x}^{c}}\left(
\mathbf{1}_{A_{x}^{c}}x+\mathbf{1}_{A_{x}}x_{0}\right)  \in\mathbf{1}%
_{A_{x}^{c}}\operatorname*{dom}f.$ Now since $f$ and the conditional
expectation operator are local, we have that%
\begin{align*}
\mathbf{1}_{A_{x}^{c}}[f(x)-\mathbb{E}_{\mu}[xy|\mathcal{G}]]  &
=\mathbf{1}_{A_{x}^{c}}[f\left(  \mathbf{1}_{A_{x}^{c}}x\right)
-\mathbb{E}_{\mu}[\mathbf{1}_{A_{x}^{c}}xy|\mathcal{G}]]\\
&  \leq\mathbf{1}_{A_{x}^{c}}\sup_{z\in\mathbf{1}_{A_{x}^{c}}%
\operatorname*{dom}f}\left\{  f\left(  z\right)  -\mathbb{E}_{\mu
}[zy|\mathcal{G}]\right\} \\
&  =\mathbf{1}_{A_{x}^{c}}\sup_{w\in\operatorname*{dom}f}\left\{  f\left(
\mathbf{1}_{A_{x}^{c}}w\right)  -\mathbb{E}_{\mu}[\mathbf{1}_{A_{x}^{c}%
}wy|\mathcal{G}]\right\} \\
&  =\mathbf{1}_{A_{x}^{c}}\sup_{w\in\operatorname*{dom}f}\left\{
\mathbf{1}_{A_{x}^{c}}f\left(  \mathbf{1}_{A_{x}^{c}}w\right)  -\mathbf{1}%
_{A_{x}^{c}}\mathbb{E}_{\mu}[\mathbf{1}_{A_{x}^{c}}wy|\mathcal{G}]\right\} \\
&  =\mathbf{1}_{A_{x}^{c}}\sup_{w\in\operatorname*{dom}f}\left\{
\mathbf{1}_{A_{x}^{c}}f\left(  w\right)  -\mathbf{1}_{A_{x}^{c}}%
\mathbb{E}_{\mu}[wy|\mathcal{G}]\right\} \\
&  =\mathbf{1}_{A_{x}^{c}}\sup_{w\in\operatorname*{dom}f}\left\{  f\left(
w\right)  -\mathbb{E}_{\mu}[wy|\mathcal{G}]\right\}  .
\end{align*}
In conclusion, since $x$ was arbitrarily chosen, we have that%
\[
c_{f}(y)=\sup_{w\in L^{\infty}(\mathcal{F}|\mathcal{G})}\left\{  f\left(
w\right)  -\mathbb{E}_{\mu}[wy|\mathcal{G}]\right\}  \leq\sup_{w\in
\operatorname*{dom}f}\left\{  f(w)-\mathbb{E}_{\mu}[wy|\mathcal{G}]\right\}
.
\]
Thus the first equality in the statement follows suit. Now we pass to the
second one. For all $y\in L^{1}(\mathcal{F}|\mathcal{G}),$ and $x_{0}%
\in\operatorname*{dom}f,$%
\[
c_{f}(y)\geq f(x_{0})-\mathbb{E}_{\mu}[x_{0}y|\mathcal{G}]>-\infty
\]
and
\[
(-f)^{\ast}(y)=\sup_{w\in L^{\infty}(\mathcal{F}|\mathcal{G})}\left\{
f(w)+\mathbb{E}_{\mu}[wy|\mathcal{G}]\right\}  \geq f(x_{0})+\mathbb{E}_{\mu
}[x_{0}y|\mathcal{G}]>-\infty
\]
since, $f(x_{0})>-\infty.$ Moreover,\ by the fact that $L^{\infty}%
(\mathcal{F}|\mathcal{G})$ satisfies the countable concatenation property and
$f$ is assumed to be proper, $L^{0}(\mathcal{G})$-concave, and$\ \left\Vert
\cdot\right\Vert _{\infty}^{\mathcal{G}}$-continuous, Corollary 3.42, and
Theorem 3.45 in \cite{Guo}, yield that $\partial(-f)(x_{0})\neq\emptyset$,
since $x_{0}\in\mathtt{int}_{\left\Vert \cdot\right\Vert _{\infty
}^{\mathcal{G}}}(\operatorname*{dom}f)$. Thus, we have%
\[
f(x_{0})+\mathbb{E}_{\mu}[x_{0}z|\mathcal{G}]=(-f)^{\ast}\left(  z\right)  .
\]
for each $z\in\partial(-f)(x_{0}).$ Since $f(x_{0})\in L^{0}(\mathcal{G}),$
the latter equality and the fact that $\partial(-f)(x_{0})\neq\emptyset$ imply
that there exists $y_{0}\in L^{1}(\mathcal{F}|\mathcal{G})$ such that
$(-f)^{\ast}\left(  y_{0}\right)  <\infty.$ Then, this yields that,%
\[
c_{f}(-y_{0})=\sup_{w\in L^{\infty}(\mathcal{F}|\mathcal{G})}\left\{  f\left(
w\right)  +\mathbb{E}_{\mu}[wy_{0}|\mathcal{G}]\right\}  =(-f)^{\ast}%
(y_{0})<\infty.
\]
Now, let $z_{0}=-y_{0},$ we proceed as in the first part of the proof. In
particular, let $x\in L^{\infty}(\mathcal{F}|\mathcal{G}).$ If $w\in
\operatorname*{dom}c_{f},$ then,
\[
\mathbb{E}_{\mu}[xw|\mathcal{G}]+c_{f}\left(  w\right)  \geq\inf
_{y\in\operatorname*{dom}c_{f}}\left\{  \mathbb{E}_{\mu}[xy|\mathcal{G}%
]+c_{f}\left(  y\right)  \right\}  .
\]
If $w\notin\operatorname*{dom}c_{f}$ then, since $c_{f}>-\infty,$ it is enough
to consider $A_{w}=\{c_{f}(w)=\infty\}.$ In particular, $A_{w}$ is a
$\mathcal{G}$-measurable set with $\mu(A_{w})>0.$ Then, we have%
\begin{align*}
\mathbf{1}_{A_{w}}[\mathbb{E}_{\mu}[xw|\mathcal{G}]+c_{f}\left(  w\right)  ]
&  =\mathbf{1}_{A_{w}}\cdot\infty\\
&  \geq\mathbf{1}_{A_{w}}\inf_{y\in\operatorname*{dom}c_{f}}\left\{
\mathbb{E}_{\mu}[xy|\mathcal{G}]+c_{f}\left(  y\right)  \right\}  \text{.}%
\end{align*}
By Lemmas \ref{scaloutfromsup} and \ref{regularity}, $c_{f}$\ is regular.
Thus, we have that%
\[
c_{f}(\mathbf{1}_{A_{w}^{c}}w+\mathbf{1}_{A_{w}}z_{0})=\mathbf{1}_{A_{w}^{c}%
}c_{f}(w)+\mathbf{1}_{A_{w}}c_{f}(z_{0})
\]
and hence since $z_{0}\in\operatorname*{dom}c_{f}$, $\mathbf{1}_{A_{w}^{c}%
}c_{f}(w)<\infty,$ and $c_{f}$ is proper, it follows
\[
\mathbf{1}_{A_{w}^{c}}w+\mathbf{1}_{A_{w}}z_{0}\in\operatorname*{dom}c_{f}.
\]
This yields $\mathbf{1}_{A_{w}^{c}}w=\mathbf{1}_{A_{w}^{c}}\left(
\mathbf{1}_{A_{w}^{c}}w+\mathbf{1}_{A_{w}}z_{0}\right)  \in\mathbf{1}%
_{A_{w}^{c}}\operatorname*{dom}c_{f}.$ Now since $c_{f}$\ and the conditional
expectation operator are local, we have that%
\begin{align*}
\mathbf{1}_{A_{w}^{c}}[\mathbb{E}_{\mu}[xw|\mathcal{G}]+c_{f}\left(  w\right)
]  &  =\mathbf{1}_{A_{w}^{c}}[\mathbb{E}_{\mu}[\mathbf{1}_{A_{w}^{c}%
}xw|\mathcal{G}]+c_{f}\left(  \mathbf{1}_{A_{w}^{c}}w\right)  ]\\
&  \geq\mathbf{1}_{A_{w}^{c}}\inf_{z\in\mathbf{1}_{A_{w}^{c}}%
\operatorname*{dom}c_{f}}\left\{  \mathbb{E}_{\mu}[xz|\mathcal{G}%
]+c_{f}\left(  z\right)  \right\} \\
&  =\mathbf{1}_{A_{w}^{c}}\inf_{y\in\operatorname*{dom}c_{f}}\left\{
\mathbb{E}_{\mu}[\mathbf{1}_{A_{w}^{c}}xy|\mathcal{G}]+c_{f}\left(
\mathbf{1}_{A_{w}^{c}}y\right)  \right\} \\
&  =\mathbf{1}_{A_{w}^{c}}\inf_{y\in\operatorname*{dom}c_{f}}\left\{
\mathbf{1}_{A_{w}^{c}}\mathbb{E}_{\mu}[\mathbf{1}_{A_{w}^{c}}xy|\mathcal{G}%
]+\mathbf{1}_{A_{w}^{c}}c_{f}\left(  \mathbf{1}_{A_{w}^{c}}y\right)  \right\}
\\
&  =\mathbf{1}_{A_{w}^{c}}\inf_{y\in\operatorname*{dom}c_{f}}\left\{
\mathbf{1}_{A_{w}^{c}}\mathbb{E}_{\mu}[xy|\mathcal{G}]+\mathbf{1}_{A_{w}^{c}%
}c_{f}\left(  y\right)  \right\} \\
&  =\mathbf{1}_{A_{w}^{c}}\inf_{y\in\operatorname*{dom}c_{f}}\left\{
\mathbb{E}_{\mu}[xy|\mathcal{G}]+c_{f}\left(  y\right)  \right\}  .
\end{align*}
Now notice that since $L^{\infty}(\mathcal{F}|\mathcal{G})$ satisfies the
countable concatenation property Proposition 2.2 and Corollary 2.2 in Zapata
\cite{Zapata1}\footnote{Notice that [18] uses the term \textit{stable }in
place of \textit{countable concatenation property}, the meaning is the same.}
yield, that $f$ is also $\sigma$-upper semicontinuous. Indeed, let
$\sigma^{\prime}=\sigma(L^{\infty}(\mathcal{F}|\mathcal{G}),\mathtt{Hom}%
_{L^{0}(\mathcal{G})}^{\left\Vert \cdot\right\Vert _{\infty}^{\mathcal{G}}%
}(L^{\infty}(\mathcal{F}|\mathcal{G}),L^{0}(\mathcal{G})))$, Zapata's results
imply that $f$ is $\sigma^{\prime}$-upper semicontinuous. Since $\left\langle
\cdot,y\right\rangle ^{\mathcal{G}}:x\mapsto\mathbb{E}_{\mu}[xy|\mathcal{G}]$
belongs to $\mathtt{Hom}_{L^{0}(\mathcal{G})}^{\left\Vert \cdot\right\Vert
_{\infty}^{\mathcal{G}}}(L^{\infty}(\mathcal{F}|\mathcal{G}),L^{0}%
(\mathcal{G}))$ for all $y\in L^{1}(\mathcal{F}|\mathcal{G})$, if $x_{\alpha
}\overset{\sigma^{\prime}}{\longrightarrow}x,$ then $x_{\alpha}\overset
{\sigma}{\longrightarrow}x.$ Therefore all $\sigma^{\prime}$-closed sets are
also $\sigma$-closed, yielding $\sigma$-upper semicontinuity of $f$. In
conclusion, since $w$ was arbitrarily chosen, by Proposition
\ref{modu-dua-rep}
\[
f(x)=\inf_{y\in L^{1}(\mathcal{F}|\mathcal{G})}\left\{  \mathbb{E}_{\mu
}[xy|\mathcal{G}]+c_{f}\left(  y\right)  \right\}  \geq\inf_{y\in
\operatorname*{dom}c_{f}}\left\{  \mathbb{E}_{\mu}[xy|\mathcal{G}%
]+c_{f}\left(  y\right)  \right\}
\]
Thus the second equality follows as well.\hfill$\blacksquare$

Now we define the following sets%
\[
M_{+}^{\circ}=\{y\in L^{1}(\mathcal{F}|\mathcal{G}):y\geq\mathbf{0}%
_{S}\}\text{ and }C=\{y\in L^{1}(\mathcal{F}|\mathcal{G}):\mathbb{E}_{\mu
}[y|\mathcal{G}]=\mathbf{1}_{S}\}.
\]
The next Lemma is crucial for the representation results we provide in the
main text and it is a small adaptation from Lemma 3.13 in \cite{AppCondRisk}
and Theorem 4.17 in \cite{Guo}.

\begin{lemma}
\label{represlemma}Let $f:L^{\infty}(\mathcal{F}|\mathcal{G})\rightarrow
\bar{L}^{0}(\mathcal{G})$ be a proper, $L^{0}(\mathcal{G})$-concave,
$\left\Vert \cdot\right\Vert _{\infty}^{\mathcal{G}}$-continuous function with
nonempty $\mathtt{int}_{\left\Vert \cdot\right\Vert _{\infty}^{\mathcal{G}}%
}(\operatorname*{dom}f).$

1. If $f$ is monotone, then for all $x\in L^{\infty}(\mathcal{F}|\mathcal{G})$%
\[
f(x)=\inf_{y\in M_{+}^{\circ}}\left\{  \mathbb{E}_{\mu}[xy|\mathcal{G}%
]+c_{f}\left(  y\right)  \right\}  .
\]
2. If $f$ is $L^{0}(\mathcal{G})$-translation invariant, then for all $x\in
L^{\infty}(\mathcal{F}|\mathcal{G})$%
\[
f(x)=\inf_{y\in C}\left\{  \mathbb{E}_{\mu}[xy|\mathcal{G}]+c_{f}\left(
y\right)  \right\}  .
\]
3. If $f$ is monotone and $L^{0}(\mathcal{G})$-translation invariant, then for
all $x\in L^{\infty}(\mathcal{F}|\mathcal{G})$%
\[
f(x)=\inf_{y\in\Delta(\mathcal{F}|\mathcal{G})}\left\{  \mathbb{E}_{\mu
}[xy|\mathcal{G}]+c_{f}\left(  y\right)  \right\}  .
\]

\end{lemma}

\noindent\textbf{Proof } Let $x_{0}\in\mathtt{int}_{\left\Vert \cdot
\right\Vert _{\infty}^{\mathcal{G}}}(\operatorname*{dom}f).$

1. Now suppose that some $z\in\operatorname*{dom}c_{f}$ and $z\notin
M_{+}^{\circ}.$ Then, we have that $\mu(\{z<0\})>0,$ and since $f$ is
monotone, for all $n\in\mathbb{N},$ we have $f(x_{0}+n\mathbf{1}%
_{\{z<0\}})\geq f(x_{0}),$ and
\begin{align*}
c_{f}(z)  &  \geq f(x_{0}+n\mathbf{1}_{\{z>0\}})-\mathbb{E}_{\mu}%
[(x_{0}+n\mathbf{1}_{\{z<0\}})z|\mathcal{G}]\\
&  \geq f(x_{0})-\mathbb{E}_{\mu}[(x_{0}+n\mathbf{1}_{\{z<0\}})z|\mathcal{G}].
\end{align*}
Therefore,%
\[
c_{f}(z)(s)\geq\left\{
\begin{array}
[c]{cc}%
+\infty & \text{\emph{if }}s\in\{z>0\}\\
0 & \text{\emph{otherwise}}%
\end{array}
\right.
\]
for all $s\in S$, contradicting the fact that $z\in\operatorname*{dom}c_{f}.$
Thus, $\operatorname*{dom}c_{f}\subseteq M_{+}^{\circ}$ and, by Lemma
\ref{domina proper}
\begin{align*}
f(x)  &  =\inf_{y\in\operatorname*{dom}c_{f}}\left\{  \mathbb{E}_{\mu
}[xy|\mathcal{G}]+c_{f}\left(  y\right)  \right\} \\
&  \geq\inf_{y\in M_{+}^{\circ}}\left\{  \mathbb{E}_{\mu}[xy|\mathcal{G}%
]+c_{f}\left(  y\right)  \right\} \\
&  \geq\inf_{y\in L^{1}(\mathcal{F}|\mathcal{G})}\left\{  \mathbb{E}_{\mu
}[xy|\mathcal{G}]+c_{f}\left(  y\right)  \right\} \\
&  =\inf_{y\in\operatorname*{dom}c_{f}}\left\{  \mathbb{E}_{\mu}%
[xy|\mathcal{G}]+c_{f}\left(  y\right)  \right\}  .
\end{align*}
for all $x\in L^{\infty}(\mathcal{F}|\mathcal{G}).$

2. Now suppose that some $z\in\operatorname*{dom}c_{f}$ and $z\notin C.$ Then,
we have that $\mu(\{\mathbb{E}_{\mu}[z|\mathcal{G}]\neq\mathbf{1}_{S}\})>0.$
Since $f$ is $L^{0}(\mathcal{G})$-translation invariant, for all $a\in
L^{0}(\mathcal{G}),$ we have%
\[
c_{f}(z)\geq f(x_{0}+a)-\mathbb{E}_{\mu}[(x_{0}+a)z|\mathcal{G}]=f(x_{0}%
)-a(\mathbb{E}_{\mu}[z|\mathcal{G}]-\mathbf{1}_{S})-\mathbb{E}_{\mu}%
[x_{0}z|\mathcal{G}].
\]
Since, $\{\mathbb{E}_{\mu}[y|\mathcal{G}]-\mathbf{1}_{S}\neq\mathbf{0}%
_{S}\}\in\mathcal{G},$ we can define the following sequence $(a_{n}%
)_{n\in\mathbb{N}}\in L^{0}(\mathcal{G})^{\mathbb{N}}$%
\[
a_{n}(s)=\left\{
\begin{array}
[c]{cc}%
-n & \text{\emph{if }}s\in\{\mathbb{E}_{\mu}[z|\mathcal{G}]-\mathbf{1}%
_{S}>0\}\\
n & \text{\emph{otherwise}}%
\end{array}
\right.
\]
then, since $\mu(\{\mathbb{E}_{\mu}[z|\mathcal{G}]\neq\mathbf{1}_{S}\})>0,$ we
have, for all $n\in\mathbb{N},$%
\[
c_{f}(z)\geq\sup_{n\in\mathbb{N}}\{f(x_{0})-a_{n}(\mathbb{E}_{\mu
}[z|\mathcal{G}]-\mathbf{1}_{S})-\mathbb{E}_{\mu}[x_{0}z|\mathcal{G}]\}.
\]
Therefore,%
\[
c_{f}(z)(s)\geq\left\{
\begin{array}
[c]{cc}%
+\infty & \text{\emph{if }}s\in\{\mathbb{E}_{\mu}[z|\mathcal{G}]\neq
\mathbf{1}_{S}\}\\
0 & \text{\emph{otherwise}}%
\end{array}
\right.
\]
contradicting $z\in\operatorname*{dom}c_{f}.$ Therefore, $\operatorname*{dom}%
c_{f}\subseteq C.$ To conclude, Lemma \ref{domina proper} and the latter
inclusion yield%
\begin{align*}
f(x)  &  =\inf_{y\in\operatorname*{dom}c_{f}}\left\{  \mathbb{E}_{\mu
}[xy|\mathcal{G}]+c_{f}\left(  y\right)  \right\} \\
&  \geq\inf_{y\in C}\left\{  \mathbb{E}_{\mu}[xy|\mathcal{G}]+c_{f}\left(
y\right)  \right\} \\
&  \geq\inf_{y\in L^{1}(\mathcal{F}|\mathcal{G})}\left\{  \mathbb{E}_{\mu
}[xy|\mathcal{G}]+c_{f}\left(  y\right)  \right\} \\
&  =\inf_{y\in\operatorname*{dom}c_{f}}\left\{  \mathbb{E}_{\mu}%
[xy|\mathcal{G}]+c_{f}\left(  y\right)  \right\}  .
\end{align*}
for all $x\in L^{\infty}(\mathcal{F}|\mathcal{G}).$

3. It follows from points 1. and 2.\hfill$\blacksquare$

\subsection{Representation result for divergences}

Fix $\phi\in\Phi$ and define $I_{\phi}:L^{\infty}(\mathcal{F}|\mathcal{G}%
)\rightarrow\bar{L}^{0}(\mathcal{G})$ as%
\[
I_{\phi}(x)=-\mathbb{E}_{\mu}[\phi^{\ast}\left(  -x\right)  |\mathcal{G}]
\]

for all $x\in L^{\infty}(\mathcal{F}|\mathcal{G}).$

\begin{lemma}
\label{PropertiesIphi}Let $\phi\in\Phi,$ then $I_{\phi}$ is $L^{0}%
(\mathcal{G})$-concave, monotone, $L^{0}(\mathcal{G})$-valued, and $\sigma
$-upper semicontinuous.
\end{lemma}

\noindent\textbf{Proof }Let $x,y\in L^{\infty}(\mathcal{F}|\mathcal{G})$ and
$a\in L^{0}(\mathcal{G})_{+}$ with $\mathbf{0}_{s}\leq a\leq\mathbf{1}_{S}$.
Then,%
\begin{align*}
aI_{\phi}(x)+(\mathbf{1}_{s}-a)I_{\phi}(y)  &  =-\mathbb{E}_{\mu}[a\phi^{\ast
}\left(  -x\right)  |\mathcal{G}]-\mathbb{E}_{\mu}[(\mathbf{1}_{S}%
-a)\phi^{\ast}\left(  -y\right)  |\mathcal{G}]\\
&  =-\mathbb{E}_{\mu}\left[  a\sup_{k\geq0}\left\{  (-x)k-\phi(k)\right\}
|\mathcal{G}\right]  -\mathbb{E}_{\mu}\left[  (\mathbf{1}_{S}-a)\sup
_{k^{\prime}\geq0}\left\{  (-y)k^{\prime}-\phi(k^{\prime})\right\}
|\mathcal{G}\right] \\
&  =-\mathbb{E}_{\mu}\left[  \sup_{k\geq0}\left\{  a(-x)k-a\phi(k)\right\}
+\sup_{k^{\prime}\geq0}\left\{  (\mathbf{1}_{S}-a)(-y)k^{\prime}%
-(\mathbf{1}_{S}-a)\phi(k^{\prime})\right\}  |\mathcal{G}\right] \\
&  \leq-\mathbb{E}_{\mu}\left[  \sup_{t\geq0}\left\{  a(-x)t-a\phi
(t)+(\mathbf{1}_{S}-a)(-y)t-(\mathbf{1}_{S}-a)\phi(t)\right\}  |\mathcal{G}%
\right] \\
&  =-\mathbb{E}_{\mu}\left[  \sup_{t\geq0}\left\{  a(-x)t+(\mathbf{1}%
_{S}-a)(-y)t-\phi(t)\right\}  |\mathcal{G}\right] \\
&  =-\mathbb{E}_{\mu}\left[  \phi^{\ast}(a(-x)+(\mathbf{1}_{S}%
-a)(-y))|\mathcal{G}\right] \\
&  =I_{\phi}(ax+(\mathbf{1}_{S}-a)y).
\end{align*}
Therefore, $I_{\phi}$ is $L^{0}(\mathcal{G})$-concave$.$

Suppose $x,y\in L^{\infty}(\mathcal{F}|\mathcal{G})$ and $x\geq y,$ since
$\phi\in\Phi$, we have that $\phi^{\ast}$ is increasing. Thus, $\phi^{\ast
}(-y)\geq\phi^{\ast}(-x),$ which implies%
\[
I_{\phi}(x)=-\mathbb{E}_{\mu}[\phi^{\ast}\left(  -x\right)  |\mathcal{G}%
]\geq-\mathbb{E}_{\mu}[\phi^{\ast}\left(  -y\right)  |\mathcal{G}]=I_{\phi
}(y).
\]
This proves that $I_{\phi}$ is monotone.

Since $\phi\in\Phi$, it follows $\phi^{\ast}(m)=\sup_{t\geq0}\left\{
mt-\phi(t)\right\}  ,$ and
\begin{align*}
-\phi^{\ast}(-m)  &  =-\sup_{t\geq0}\left\{  -mt-\phi(t)\right\} \\
&  =\inf_{t\geq0}\left\{  mt+\phi(t)\right\} \\
&  \leq m.
\end{align*}
for all $m\in\mathbb{R}$. This yields that, for all $x\in L^{\infty
}(\mathcal{F}|\mathcal{G}),$%
\[
I_{\phi}(x)=\mathbb{E}_{\mu}[-\phi^{\ast}\left(  -x\right)  |\mathcal{G}%
]\leq\mathbb{E}_{\mu}[x|\mathcal{G}]\in L^{0}(\mathcal{G}).
\]
Now fix, $x\in L^{\infty}(\mathcal{F}|\mathcal{G})$, then $x\geq-|x|,$
therefore $-\phi^{\ast}\left(  -x\right)  \geq-\phi^{\ast}\left(
-(-|x|)\right)  =-\phi^{\ast}\left(  |x|\right)  .$ Thus,%
\begin{align*}
I_{\phi}(x)  &  =\mathbb{E}_{\mu}[-\phi^{\ast}\left(  -x\right)
|\mathcal{G}]\\
&  \geq\mathbb{E}_{\mu}[-\phi^{\ast}\left(  |x|\right)  |\mathcal{G}]\\
&  \geq\mathbb{E}_{\mu}\left[  -\phi^{\ast}\left(  \left\Vert x\right\Vert
_{\infty}^{\mathcal{G}}\right)  |\mathcal{G}\right] \\
&  =-\phi^{\ast}\left(  \left\Vert x\right\Vert _{\infty}^{\mathcal{G}%
}\right)  \in L^{0}(\mathcal{G})
\end{align*}
which yields that $I_{\phi}$ is $L^{0}(\mathcal{G})$-valued.

To conclude, suppose $(x_{n})_{n\in\mathbb{N}}$ is a sequence in $L^{\infty
}(\mathcal{F}|\mathcal{G})$ such that $\left\Vert x_{n}-x\right\Vert _{\infty
}^{\mathcal{G}}\rightarrow\mathbf{0}_{S}$ for some $x\in L^{\infty
}(\mathcal{F}|\mathcal{G}).$ Then, it follows that $\mathbf{0}_{S}%
\leq\left\vert x_{n}-x\right\vert \leq\left\Vert x_{n}-x\right\Vert _{\infty
}^{\mathcal{G}}\rightarrow\mathbf{0}_{S}$ thus $x_{n}$ converges to $x$ with
respect to the strong order topology. Since $\phi^{\ast}$ is convex and
increasing, it is also continuous, therefore $\phi^{\ast}(-x_{n})$ converges
to $\phi^{\ast}(-x)$ in the strong order topology. Then, by the continuity of
the conditional expectation operator,%
\[
\mathbf{0}_{S}\leq\left\vert I_{\phi}(x_{n})-I_{\phi}(x)\right\vert
\leq\mathbb{E}_{\mu}[\left\vert \phi^{\ast}\left(  -x_{n}\right)  -\phi^{\ast
}\left(  -x\right)  \right\vert |\mathcal{G}]\rightarrow\mathbf{0}_{S}%
\]
proving that $I_{\phi}$ is $\left\Vert \cdot\right\Vert _{\infty}%
^{\mathcal{G}}$-continuous, and hence, by Corollary 2.2 in \cite{Zapata1},
$I_{\phi}$ is $\sigma^{\prime}$-upper semicontinuous,\footnote{Recall that%
\[
\sigma^{\prime}=\sigma\left(  L^{\infty}(\mathcal{F}|\mathcal{G}%
),\mathtt{Hom}_{L^{0}(\mathcal{G})}^{\left\Vert \cdot\right\Vert _{\infty
}^{\mathcal{G}}}(L^{\infty}(\mathcal{F}|\mathcal{G}),L^{0}(\mathcal{G}%
))\right)  .
\]
} and thus, it is also $\sigma$-upper semicontinuous. Indeed, $\left\langle
\cdot,y\right\rangle ^{\mathcal{G}}:x\mapsto\mathbb{E}_{\mu}[xy|\mathcal{G}]$
belong to $\mathtt{Hom}_{L^{0}(\mathcal{G})}^{\left\Vert \cdot\right\Vert
_{\infty}^{\mathcal{G}}}(L^{\infty}(\mathcal{F}|\mathcal{G}),L^{0}%
(\mathcal{G}))$ for all $y\in L^{1}(\mathcal{F}|\mathcal{G})$, and hence, if
$x_{\alpha}\overset{\sigma^{\prime}}{\longrightarrow}x,$ then $x_{\alpha
}\overset{\sigma}{\longrightarrow}x,$ therefore all $\sigma^{^{\prime}}%
$-closed sets are also $\sigma$-closed, yielding $\sigma$-upper
semicontinuity.\hfill$\blacksquare$

We denote by $\Delta^{\sigma}\left(  \mathcal{F}\right)  $ the set of
countably additive probability measures on $\mathcal{F}$.

\begin{lemma}
\label{changemeas}Let $x\in L^{\infty}(\mathcal{F}|\mathcal{G})$ and $\nu
\in\Delta^{\sigma}\left(  \mathcal{F}\right)  $ with$\ \nu\ll\mu$, then%
\[
\mathbb{E}_{\mu}\left[  x\frac{d\nu}{d\mu}{\LARGE |}\mathcal{G}\right]
=\mathbb{E}_{\nu}\left[  x{\LARGE |}\mathcal{G}\right]  \mathbb{E}_{\mu
}\left[  \frac{d\nu}{d\mu}{\LARGE |}\mathcal{G}\right]  .
\]

\end{lemma}

\noindent\textbf{Proof }Let $B\in\mathcal{G},$ $x\in L^{\infty}(\mathcal{F}%
|\mathcal{G})_{+},$ and $n\in\mathbb{N}$,
\[
\int_{B}\mathbb{E}_{\mu}\left[  (x\wedge n)\frac{d\nu}{d\mu}{\Huge |}%
\mathcal{G}\right]  d\mu=\int_{B}(x\wedge n)\frac{d\nu}{d\mu}d\mu=\int
_{B}(x\wedge n)d\nu.
\]
Now, since $\mathbb{E}_{\nu}\left[  x\wedge n{\LARGE |}\mathcal{G}\right]  $
is $\mathcal{G}$-measurable, we have that%
\[
\mathbb{E}_{\mu}\left[  \mathbb{E}_{\nu}\left[  x\wedge n{\LARGE |}%
\mathcal{G}\right]  \frac{d\nu}{d\mu}{\LARGE |}\mathcal{G}\right]
=\mathbb{E}_{\nu}\left[  x\wedge n{\LARGE |}\mathcal{G}\right]  \mathbb{E}%
_{\mu}\left[  \frac{d\nu}{d\mu}{\LARGE |}\mathcal{G}\right]
\]
and integrating over $B$ with respect to $\mu,$%
\begin{align*}
\int_{B}\mathbb{E}_{\nu}\left[  x\wedge n{\LARGE |}\mathcal{G}\right]
\mathbb{E}_{\mu}\left[  \frac{d\nu}{d\mu}{\LARGE |}\mathcal{G}\right]  d\mu &
=\int_{B}\mathbb{E}_{\mu}\left[  \mathbb{E}_{\nu}\left[  x\wedge
n{\LARGE |}\mathcal{G}\right]  \frac{d\nu}{d\mu}{\LARGE |}\mathcal{G}\right]
d\mu\\
&  =\int_{B}\mathbb{E}_{\nu}\left[  x\wedge n{\LARGE |}\mathcal{G}\right]
\frac{d\nu}{d\mu}d\mu\\
&  =\int_{B}\mathbb{E}_{\nu}\left[  x\wedge n{\LARGE |}\mathcal{G}\right]
d\nu\\
&  =\int_{B}(x\wedge n)d\nu\\
&  =\int_{B}\mathbb{E}_{\mu}\left[  (x\wedge n)\frac{d\nu}{d\mu}%
{\LARGE |}\mathcal{G}\right]  d\mu.
\end{align*}
Thus, since $B$ was chosen arbitrarily we have that
\[
\mathbb{E}_{\mu}\left[  (x\wedge n)\frac{d\nu}{d\mu}{\LARGE |}\mathcal{G}%
\right]  =\mathbb{E}_{\nu}\left[  x\wedge n{\LARGE |}\mathcal{G}\right]
\mathbb{E}_{\mu}\left[  \frac{d\nu}{d\mu}{\LARGE |}\mathcal{G}\right]  .
\]
Since this equality holds for each $n\in\mathbb{N},$ we have that%
\begin{align*}
\mathbb{E}_{\mu}\left[  x\frac{d\nu}{d\mu}{\LARGE |}\mathcal{G}\right]   &
=\lim_{n\rightarrow\infty}\mathbb{E}_{\mu}\left[  x\frac{d\nu}{d\mu}\wedge
n{\LARGE |}\mathcal{G}\right] \\
&  =\lim_{n\rightarrow\infty}\mathbb{E}_{\mu}\left[  (x\wedge n)\frac{d\nu
}{d\mu}{\LARGE |}\mathcal{G}\right] \\
&  =\lim_{n\rightarrow\infty}\mathbb{E}_{\nu}\left[  x\wedge n{\LARGE |}%
\mathcal{G}\right]  \mathbb{E}_{\mu}\left[  \frac{d\nu}{d\mu}{\LARGE |}%
\mathcal{G}\right] \\
&  =\mathbb{E}_{\nu}\left[  x{\LARGE |}\mathcal{G}\right]  \mathbb{E}_{\mu
}\left[  \frac{d\nu}{d\mu}{\LARGE |}\mathcal{G}\right]
\end{align*}
where the second equality follows from the fact that $\frac{d\nu}{d\mu}$ is a
nonnegative function integrating to 1, and we are considering the limit for
$n\rightarrow\infty$. Now let $x\in L^{\infty}(\mathcal{F}|\mathcal{G}),$ then%
\begin{align*}
\mathbb{E}_{\mu}\left[  x\frac{d\nu}{d\mu}{\LARGE |}\mathcal{G}\right]   &
=\mathbb{E}_{\mu}\left[  (x^{+}-x^{-})\frac{d\nu}{d\mu}{\LARGE |}%
\mathcal{G}\right] \\
&  =\mathbb{E}_{\mu}\left[  x^{+}\frac{d\nu}{d\mu}{\LARGE |}\mathcal{G}%
\right]  -\mathbb{E}_{\mu}\left[  x^{-}\frac{d\nu}{d\mu}{\LARGE |}%
\mathcal{G}\right] \\
&  =\mathbb{E}_{\nu}\left[  x^{+}{\LARGE |}\mathcal{G}\right]  \mathbb{E}%
_{\mu}\left[  \frac{d\nu}{d\mu}{\LARGE |}\mathcal{G}\right]  -\mathbb{E}_{\nu
}\left[  x^{-}{\LARGE |}\mathcal{G}\right]  \mathbb{E}_{\mu}\left[  \frac
{d\nu}{d\mu}{\LARGE |}\mathcal{G}\right] \\
&  =\mathbb{E}_{\nu}\left[  x{\LARGE |}\mathcal{G}\right]  \mathbb{E}_{\mu
}\left[  \frac{d\nu}{d\mu}{\LARGE |}\mathcal{G}\right]  .
\end{align*}
$\blacksquare$

Here we adopt the notation presented in section 4. For some $\phi\in\Phi$ and
$\hat{\phi}\in C\left(  \phi\right)  $ we define $J_{\hat{\phi}}:L^{\infty
}(\mathcal{F}|\mathcal{G})\rightarrow\bar{L}^{0}(\mathcal{G})$ as%
\[
J_{\hat{\phi}}(x)=\mathbb{E}_{\mu}[\hat{\phi}(x)|\mathcal{G}]
\]
for all $x\in L^{\infty}(\mathcal{F}|\mathcal{G}).$ Given, the properties of
$\hat{\phi},$ we have that $J_{\hat{\phi}}$ is $L^{0}(\mathcal{G})-$convex and
well defined.\footnote{$L^{0}(\mathcal{G})$-convexity follows the same steps
we used above, and therefore we redirect to the proof of Lemma 12. It is also
immediate to see that for all $\hat{\phi}\in C\left(  \phi\right)  $, it
follows $\hat{\phi}(m)\geq mt-\phi(t)$ for all $m\in\mathbb{R}$ and all
$t\geq0.$ Thus, $\hat{\phi}(m)\geq m$ for all $m\in\mathbb{R}$. This yields,
\[
J_{\hat{\phi}}(x)=\mathbb{E}_{\mu}\left[  \hat{\phi}(x)|\mathcal{G}\right]
\geq\mathbb{E}_{\mu}\left[  x|\mathcal{G}\right]  \in L^{0}(\mathcal{G})
\]
\par
for all $x\in L^{\infty}(\mathcal{F}|\mathcal{G}).$
\par
{}} Moreover, if we restrict the domain of the convex conjugate $\left(
J_{\hat{\phi}}\right)  ^{\ast}$ to $\Delta(\mathcal{F}|\mathcal{G}),$ by Lemma
\ref{measrep} we can assume that the domain of $\left(  J_{\hat{\phi}}\right)
^{\ast}|_{\Delta(\mathcal{F}|\mathcal{G})}$ is exactly $\mathcal{M(G)}.$

\begin{theorem}
\label{thm:CondiRocky}Let $\phi\in\Phi$ and $\hat{\phi}\in C\left(
\phi\right)  $.\ If $\hat{\phi}$\ is real valued, then,%
\[
\left(  J_{\hat{\phi}}\right)  ^{\ast}|_{\Delta(\mathcal{F}|\mathcal{G}%
)}\left(  \nu\right)  =\mathbb{E}_{\mu}\left[  \phi\left(  \frac{d\nu}{d\mu
}\right)  {\LARGE |}\mathcal{G}\right]  \ \ \ \qquad\forall\nu\in
\mathcal{M(G)}%
\]

\end{theorem}

\noindent\textbf{Proof}\ Since $\hat{\phi}$ is real valued and convex,\ it
follows that $\hat{\phi}$ is continuous. Now, let $\nu\in\mathcal{M(G)},$ then
$\nu|_{\mathcal{G}}=\mu|_{\mathcal{G}}$ and $\nu\ll\mu$. Let $\frac{d\nu}%
{d\mu}$ be any nonnegative real valued $\mathcal{F}$-measurable function such
that%
\[
\nu\left(  E\right)  =\int_{E}\frac{d\nu}{d\mu}d\mu\qquad\forall
E\in\mathcal{F}.
\]
Moreover, notice that since $\nu|_{\mathcal{G}}=\mu|_{\mathcal{G}}$ we have
that
\[
\mathbb{E}_{\mu}\left[  \frac{d\nu}{d\mu}{\LARGE |}\mathcal{G}\right]
=\mathbf{1}_{S}.
\]
Indeed, both functions are $\mathcal{G}$-measurable and integrable, thus for
all $B\in$ $\mathcal{G}$,%
\[
\int_{B}\mathbb{E}_{\mu}\left[  \frac{d\nu}{d\mu}{\LARGE |}\mathcal{G}\right]
d\mu=\int_{B}\frac{d\nu}{d\mu}d\mu=\int_{B}\mathbf{1}_{S}d\nu=\nu
(B)=\mu(B)=\int_{B}\mathbf{1}_{S}d\mu.
\]
By Lemma \ref{changemeas} it follows%

\[
\mathbb{E}_{\mu}\left[  x\frac{d\nu}{d\mu}{\LARGE |}\mathcal{G}\right]
=\mathbb{E}_{\nu}\left[  x{\LARGE |}\mathcal{G}\right]  \mathbb{E}_{\mu
}\left[  \frac{d\nu}{d\mu}{\LARGE |}\mathcal{G}\right]  =\mathbb{E}_{\nu
}\left[  x{\LARGE |}\mathcal{G}\right]  .
\]
By definition of convex conjugate, we have that $mt-\hat{\phi}\left(
m\right)  \leq\hat{\phi}^{\ast}\left(  t\right)  $ for all $m,t\in\mathbb{R}$.
We have that for each $x\in L^{\infty}(\mathcal{F}|\mathcal{G})$%
\[
x\left(  s\right)  \frac{d\nu}{d\mu}\left(  s\right)  -\hat{\phi}\left(
x\left(  s\right)  \right)  \leq\hat{\phi}^{\ast}\left(  \frac{d\nu}{d\mu
}\left(  s\right)  \right)  =\phi\left(  \frac{d\nu}{d\mu}\left(  s\right)
\right)  \quad\forall s\in S.
\]
Then, by definition of $\left(  J_{\hat{\phi}}\right)  ^{\ast}$, for all
$\nu\in\mathcal{M(G)},$%
\begin{align*}
\left(  J_{\hat{\phi}}\right)  ^{\ast}|_{\Delta(\mathcal{F}|\mathcal{G}%
)}\left(  \nu\right)   &  =\sup_{x\in L^{\infty}(\mathcal{F}|\mathcal{G}%
)}\left\{  \mathbb{E}_{\nu}[x|\mathcal{G}]-\mathbb{E}_{\mu}\left[  \hat{\phi
}(x)|\mathcal{G}\right]  \right\} \\
&  =\sup_{x\in L^{\infty}(\mathcal{F}|\mathcal{G})}\left\{  \mathbb{E}_{\mu
}\left[  x\frac{d\nu}{d\mu}{\LARGE |}\mathcal{G}\right]  -\mathbb{E}_{\mu
}\left[  \hat{\phi}(x)|\mathcal{G}\right]  \right\} \\
&  =\sup_{x\in L^{\infty}(\mathcal{F}|\mathcal{G})}\left\{  \mathbb{E}_{\mu
}\left[  x\frac{d\nu}{d\mu}-\hat{\phi}(x)|\mathcal{G}\right]  \right\} \\
&  \leq\mathbb{E}_{\mu}\left[  \phi\left(  \frac{d\nu}{d\mu}\right)
{\LARGE |}\mathcal{G}\right]  .
\end{align*}
Recall that%
\[
\phi\left(  t\right)  =\hat{\phi}^{\ast}\left(  t\right)  =\sup_{m\in
\mathbb{R}}\left\{  mt-\hat{\phi}\left(  m\right)  \right\}  \quad\forall
t\geq0.
\]
Note that for each $t>0$ and for each $m\in\partial\phi\left(  t\right)  $%
\[
\phi\left(  t\right)  =mt-\hat{\phi}\left(  m\right)
\]
Recall that $\phi_{+}^{\prime}\left(  t\right)  \in\partial\phi\left(
t\right)  $ for all $t>0$. Since $\phi$ is convex and $\phi\left(  1\right)
=0$ as well as $\phi\geq0$, we have that $0\in\partial\phi\left(  1\right)  $
and $0=\phi\left(  1\right)  =0\times1-\hat{\phi}\left(  0\right)  $, yielding
that $\hat{\phi}\left(  0\right)  =0$.\ Since $\phi\left(  t\right)
=\sup_{m\in\mathbb{R}}\left\{  mt-\hat{\phi}\left(  m\right)  \right\}  $ for
all $t\geq0$, we have that $\phi\left(  0\right)  =\sup_{m\in\mathbb{R}%
}\left\{  -\hat{\phi}\left(  m\right)  \right\}  $, yielding that there exists
a sequence $(m_{n})_{n\in%
\mathbb{N}
}$ such that $-\hat{\phi}\left(  m_{n}\right)  \uparrow\phi\left(  0\right)
$. Define $f:\left[  0,\infty\right)  \rightarrow\mathbb{R}$ by%
\[
f\left(  t\right)  =\left\{
\begin{array}
[c]{cc}%
\phi_{+}^{\prime}\left(  t\right)  & t\in\left(  0,1\right)  \cup\left(
1,\infty\right) \\
0 & t=0,1
\end{array}
\right.  \qquad\forall t\in\left[  0,\infty\right)
\]
Since $\phi$ is strictly convex, we have that $\phi_{+}^{\prime}$ is strictly
increasing and\ $f$ is Borel measurable. Define
\[
B_{0}=\left\{  s\in S:\frac{d\nu}{d\mu}\left(  s\right)  =0\right\}  ,\text{
}B_{1}=\left\{  s\in S:\frac{d\nu}{d\mu}\left(  s\right)  =1\right\}  ,\text{
}B_{2}=\left\{  s\in S:1\not =\frac{d\nu}{d\mu}\left(  s\right)  >0\right\}
,
\]
Note also that $f\left(  t\right)  \in\partial\phi\left(  t\right)  $ for all
$t>0$. Since $\phi$ is strictly convex and $\phi\left(  1\right)  =0$ as well
as $\phi\geq0$, it follows that $g:S\rightarrow\mathbb{R}$ by $g=f\left(
\frac{d\nu}{d\mu}\right)  $ is $\mathcal{F}$-measurable and\footnote{At the
cost of being pedantic, suppose there exists $s^{\prime}\in B_{2}$ such that
$\phi\left(  \frac{d\nu}{d\mu}\left(  s^{\prime}\right)  \right)  =0.$ Then,
if $s\in B_{1}$,%
\[
0\leq\phi\left(  \frac{1}{2}\frac{d\nu}{d\mu}\left(  s^{\prime}\right)
+\frac{1}{2}\frac{d\nu}{d\mu}\left(  s\right)  \right)  <\frac{1}{2}%
\phi\left(  \frac{d\nu}{d\mu}\left(  s^{\prime}\right)  \right)  =0
\]
a contradiction.}%
\begin{align}
0  &  <\phi\left(  \frac{d\nu}{d\mu}\left(  s\right)  \right)  =g\left(
s\right)  \frac{d\nu}{d\mu}\left(  s\right)  -\hat{\phi}\left(  g\left(
s\right)  \right)  \qquad\forall s\in B_{2}\label{eq:str-pos}\\
0  &  =g\left(  s\right)  \frac{d\nu}{d\mu}\left(  s\right)  -\hat{\phi
}\left(  g\left(  s\right)  \right)  \qquad\forall s\in B_{0}\cup
B_{1}.\nonumber
\end{align}
Consider a sequence of simple $\mathcal{F}$-measurable random variables
$\left(  \varphi_{n}\right)  _{_{n\in%
\mathbb{N}
}}$ such that $\varphi_{n}\left(  s\right)  \rightarrow g\left(  s\right)  $
for all $s\in S$ (see, e.g., Theorem 13.5 in \cite{Billi}). For all $n\in%
\mathbb{N}
$ define $A_{n}=\{s\in S:\varphi_{n}\left(  s\right)  \frac{d\nu}{d\mu}\left(
s\right)  -\hat{\phi}\left(  \varphi_{n}\left(  s\right)  \right)  >0\}$ and
$\psi_{n}=m_{n}1_{B_{0}}+1_{A_{n}\cap B_{2}}\varphi_{n}$. Let $s\in S$. We
have three cases:

\begin{enumerate}
\item $s\in B_{0}$. It follows that $\lim_{n}\left\{  \psi_{n}\left(
s\right)  \frac{d\nu}{d\mu}\left(  s\right)  -\hat{\phi}\left(  \psi
_{n}\left(  s\right)  \right)  \right\}  =\lim_{n}\left\{  -\hat{\phi}\left(
m_{n}\right)  \right\}  =\phi\left(  0\right)  =\phi\left(  \frac{d\nu}{d\mu
}\left(  s\right)  \right)  $. Moreover, we have that%
\[
\psi_{n}\left(  s\right)  \frac{d\nu}{d\mu}\left(  s\right)  -\hat{\phi
}\left(  \psi_{n}\left(  s\right)  \right)  =-\hat{\phi}\left(  m_{n}\right)
\geq-\hat{\phi}\left(  m_{1}\right)  .
\]
for all $n\in\mathbb{N}$.

\item $s\in B_{1}$. It follows that $\psi_{n}\left(  s\right)  \frac{d\nu
}{d\mu}\left(  s\right)  -\hat{\phi}\left(  \psi_{n}\left(  s\right)  \right)
=0=\phi\left(  1\right)  =\phi\left(  \frac{d\nu}{d\mu}\left(  s\right)
\right)  $ for all $n\in%
\mathbb{N}
$. In particular, we have that $\lim_{n}\left\{  \psi_{n}\left(  s\right)
\frac{d\nu}{d\mu}\left(  s\right)  -\hat{\phi}\left(  \psi_{n}\left(
s\right)  \right)  \right\}  =\phi\left(  \frac{d\nu}{d\mu}\left(  s\right)
\right)  $. Moreover, we have that%
\[
\psi_{n}\left(  s\right)  \frac{d\nu}{d\mu}\left(  s\right)  -\hat{\phi
}\left(  \psi_{n}\left(  s\right)  \right)  \geq0.
\]
for all $n\in\mathbb{N}$.

\item $s\in B_{2}$. By (\ref{eq:str-pos})\ and since $\varphi_{n}\left(
s\right)  \rightarrow g\left(  s\right)  $ and $\hat{\phi}$ is continuous, we
have that there exists $\bar{n}\in%
\mathbb{N}
$ such that%
\[
\varphi_{n}\left(  s\right)  \frac{d\nu}{d\mu}\left(  s\right)  -\hat{\phi
}\left(  \varphi_{n}\left(  s\right)  \right)  >0\qquad\forall n\geq\bar{n}.
\]
It follows that $s\in A_{n}$ for all $n\geq\bar{n}$. This implies that
$\psi_{n}\left(  s\right)  =\varphi_{n}\left(  s\right)  $ for all $n\geq
\bar{n}$. We can conclude that%
\[
\lim_{n}\left\{  \psi_{n}\left(  s\right)  \frac{d\nu}{d\mu}\left(  s\right)
-\hat{\phi}\left(  \psi_{n}\left(  s\right)  \right)  \right\}  =\lim
_{n}\left\{  \varphi_{n}\left(  s\right)  \frac{d\nu}{d\mu}\left(  s\right)
-\hat{\phi}\left(  \varphi_{n}\left(  s\right)  \right)  \right\}
=\phi\left(  \frac{d\nu}{d\mu}\left(  s\right)  \right)  .
\]
Moreover, we have that either $s\in A_{n}$ or $s\in A_{n}^{c}$. In the first
case, since $s\in A_{n}\cap B_{2}$, we have that $\psi_{n}\left(  s\right)
=\varphi_{n}\left(  s\right)  $ and%
\[
\psi_{n}\left(  s\right)  \frac{d\nu}{d\mu}\left(  s\right)  -\hat{\phi
}\left(  \psi_{n}\left(  s\right)  \right)  =\varphi_{n}\left(  s\right)
\frac{d\nu}{d\mu}\left(  s\right)  -\hat{\phi}\left(  \varphi_{n}\left(
s\right)  \right)  >0
\]
In the second case, since $s\in A_{n}^{c}\cap B_{2}$, we have that $\psi
_{n}\left(  s\right)  =0$ and%
\[
\psi_{n}\left(  s\right)  \frac{d\nu}{d\mu}\left(  s\right)  -\hat{\phi
}\left(  \psi_{n}\left(  s\right)  \right)  =0
\]
By points 1--3 and since $S=B_{0}\cup B_{1}\cup B_{2}$ and $s$\ was
arbitrarily chosen, we have that%
\begin{equation}
\lim_{n}\left\{  \psi_{n}\left(  s\right)  \frac{d\nu}{d\mu}\left(  s\right)
-\hat{\phi}\left(  \psi_{n}\left(  s\right)  \right)  \right\}  =\phi\left(
\frac{d\nu}{d\mu}\left(  s\right)  \right)  \qquad\forall s\in S
\label{eq:app-phi-der}%
\end{equation}
and $\psi_{n}\frac{d\nu}{d\mu}-\hat{\phi}\left(  \psi_{n}\right)  \geq
\min\left\{  0,-\hat{\phi}\left(  m_{1}\right)  \right\}  $ for all $n\in%
\mathbb{N}
$. By the conditional Fatou's Lemma, we have that%
\begin{align*}
\left(  J_{\hat{\phi}}\right)  ^{\ast}|_{\Delta(\mathcal{F}|\mathcal{G}%
)}\left(  \nu\right)   &  =\sup_{x\in L^{\infty}(\mathcal{F}|\mathcal{G}%
)}\left\{  \mathbb{E}_{\nu}[x|\mathcal{G}]-\mathbb{E}_{\mu}[\hat{\phi
}(x)|\mathcal{G}]\right\}  =\sup_{x\in L^{\infty}(\mathcal{F}|\mathcal{G}%
)}\left\{  \mathbb{E}_{\mu}\left[  x\frac{d\nu}{d\mu}{\LARGE |}\mathcal{G}%
\right]  -\mathbb{E}_{\mu}\left[  \hat{\phi}(x)|\mathcal{G}\right]  \right\}
\\
&  \geq\sup_{n}\left\{  \mathbb{E}_{\mu}\left[  \psi_{n}\frac{d\nu}{d\mu
}{\LARGE |}\mathcal{G}\right]  -\mathbb{E}_{\mu}\left[  \hat{\phi}(\psi
_{n}){\LARGE |}\mathcal{G}\right]  \right\}  \geq\liminf_{n}\left\{
\mathbb{E}_{\mu}\left[  \psi_{n}\frac{d\nu}{d\mu}-\hat{\phi}(\psi
_{n}){\LARGE |}\mathcal{G}\right]  \right\} \\
&  \geq\mathbb{E}_{\mu}\left[  \phi\left(  \frac{d\nu}{d\mu}\right)
{\LARGE |}\mathcal{G}\right]
\end{align*}
proving the opposite inequality. This proves that for all $\nu\in
\mathcal{M(G)}$,
\[
\left(  J_{\hat{\phi}}\right)  ^{\ast}|_{\Delta(\mathcal{F}|\mathcal{G}%
)}\left(  \nu\right)  =\mathbb{E}_{\mu}\left[  \phi\left(  \frac{d\nu}{d\mu
}\right)  {\LARGE |}\mathcal{G}\right]  .
\]
$\blacksquare$
\end{enumerate}

\end{document}